\documentclass{aa}
\usepackage[varg]{txfonts}
\usepackage{longtable}
\usepackage{hyperref}
\hypersetup{
	colorlinks=true,
	citecolor=blue,
	urlcolor=magenta,
	linkcolor=magenta
}
\usepackage{lscape}
\usepackage{amsmath}
\usepackage{subfiles}

\usepackage{placeins}

\begin{document}

\newcommand{\rhk}{$\log R'_\text{HK}$}
\newcommand{\cahk}{\ion{Ca}{ii} H\&K}
\newcommand{\ca}{\ion{Ca}{ii}}
\newcommand{\sca}{$S_\ion{Ca}{ii}$}
\newcommand{\ha}{H$\alpha$}
\newcommand{\sha}{$S_{\mathrm{H}\alpha}$}
\newcommand{\nsha}{$S_{\mathrm{H}\alpha06}$}
\newcommand{\wsha}{$S_{\mathrm{H}\alpha16}$}

\title{Optimising the \ha~index for the identification of activity signals in FGK stars}
\subtitle{Improvement of the correlation between \ha{} and \cahk{}}

\author{
    J.~Gomes da Silva\inst{1}\thanks{\email{Joao.Silva@astro.up.pt}},
	A. Bensabat\inst{1},
	T. Monteiro\inst{1},
	N. C. Santos\inst{1,2}
    }

\institute{
    \inst{1}Instituto de Astrof\'isica e Ci\^{e}ncias do Espa\c{c}o, Porto, Portugal\\
	\inst{2}Departamento de F\'isica e Astronomia, Faculdade de Ci\^{e}ncias da Universidade do Porto, Portugal
    }

\date{Received XXX / Accepted XXX}

\abstract 
{The \cahk{} and \ha{} lines are two of the most used activity diagnostics to detect stellar activity signals in the optical regime, and to infer about possible false positives in exoplanet detection with the radial velocity method. The flux in the two lines is known to follow the solar activity cycle, and to correlate well with sunspot number and other activity diagnostics. However, for other stars, the flux in these lines is known to have a wide range of correlations, increasing the difficulty in the interpretation of the signals observed with the \ha{} line.}
{In this work we investigate the effect of the \ha{} bandpass width in the correlation between the \ca{} and \ha{} indices with the aim of improving the \ha{} index to better identify and model the signals coming from activity variability.}
{We used a sample of 152 FGK dwarfs observed with HARPS for more than 13 years with enough cadence to be able to detect rotational modulations and cycles in activity proxies.
We calculated the \ca{} and \ha{} activity indices using a range of bandwidths for \ha{} between 0.1 and 2.0 \AA{}.
We studied the correlation between the indices time series at long and short timescales and analysed the impact of stellar parameters, activity level and variability on the correlations.}
{The correlation between \ca{} and \ha{} both at short and long timespans is maximised when using narrow \ha{} bandwidths, with a maximum at 0.6 \AA{}. For some inactive stars, as the activity level increases, the flux in the \ha{} line core increases while the flux in the line wings decreases as the line becomes shallower and broader. The balance between these fluxes can cause stars to show the negative correlations observed in the literature when using a wide bandwidth on \ha{}. These anti-correlations may become positive correlations if using the 0.6 \AA{} bandwidth.
We demonstrate that rotationally modulated signals observed in \sca{}, that appear flat or noisy when using 1.6 \AA{} on \sha{}, can become more evident if a 0.6 \AA{} bandpass is used instead.
Low activity variability appears to be a contributing factor for the cases of weak or no correlations.}
{Calculating the \ha{} index using a bandpass of 0.6 \AA{} maximises the correlation between \ca{} and \ha{} both at short and long timescales. On the other hand, the use of the broader 1.6 \AA{}, generally used in exoplanet detection to identify stellar activity signals, degrades the signal by including the flux in the line wings. In face to these results we strongly recommend the use of a 0.6 \AA{} bandwidth when computing the \ha{} index for the identification of activity rotational modulation and magnetic cycle signals in solar-type stars.}

\keywords{Stars: activity -- Stars: solar-type -- Planets and satellites: detection -- Techniques: spectroscopic}

\authorrunning{J. Gomes da Silva et al.}
\maketitle

\section{Introduction}
As planet hunting instrumentation reaches higher precision levels, stellar activity becomes one of the most important limiting factors to exoplanet detection and characterization.
At short timescales of days to months, rotational modulation of active regions, zones of strong magnetic fields characterized by the presence of dark and cool spots and/or bright and hotter plages, create radial velocity (RV) quasi-periodic variations with timescales close to the rotational period, its harmonics, and that of active region life times, which can be of tens of rotational periods \citep[e.g.][]{saar1997, queloz2001, santos2000, boisse2009, boisse2011, santos2014, faria2020}.
At longer timescales of years and decades, activity variability similar to that of the 11-year solar cycle is also known to affect RV and can interfere with the detection of long-period planets \citep[e.g.][]{lovis2011, gomesdasilva2012}.

Several techniques to correct the effects of stellar activity on RV require the simultaneous measurement of activity proxies based on activity sensitive spectral lines (or using line shape indicators) to identify and remove activity effects on RV \citep[e.g.][]{boisse2011, dumusque2012, rajpaul2015, faria2022}.
The most widely used activity sensitive lines in the optical regime are the \cahk{} and \ha{} lines.
The flux on both lines is known to correlate with the presence of active regions and both are known to follow tightly the solar magnetic cycle \citep{livingston2007}.
However, studies of time series of indices based on the two lines for other stars have shown that the correlation between the indices is not straightforward: some stars show strong positive correlations, for others the two indices are not correlated, while a few show negative correlations \citep{cincunegui2007b, gomesdasilva2011, gomesdasilva2014, meunier2022}.
Although the correlation between \ca{} and \ha{} is not fully understood, some studies have found that the correlation tends to be strongly positive for stars with higher activity levels, while in the low activity regime, all types of correlations are observed \citep{walkowicz2009, gomesdasilva2011, gomesdasilva2014, meunier2022}.

Since the cores of the \cahk{} and \ha{} lines are formed at different temperatures, and thus different heights in the chromosphere \citep[e.g.][]{mauas1994, mauas1996, mauas1997, fontenla2016}, we expect these two lines to trace activity phenomena differently.
The different sensitivity of these lines to spots, plages/faculae, or filaments could explain the various correlations observed for different stars due to their varying filling factors, spatial distribution and/or contrasts \citep[e.g.][]{meunier2009}.
Although some previous works have used a narrow bandpass of 0.678 \AA{} to compute \ha{} \citep[e.g.][]{kurster2003, bonfils2007, boisse2009, santos2010}, the majority of the recent studies of stellar activity, mainly in the context of exoplanet detection, have used broader bandwidths, usually of 1.6 \AA{}.
In fact, most studies of the correlation between \ca{} and \ha{}, for FGK stars or the Sun, have also used a broad bandwidth to measure the \ha{} index \citep{cincunegui2007b, gomesdasilva2014, maldonado2019, meunier2022}.

In this work we analyse the effect of changing the \ha{} bandpass width on the correlation between the flux in the \ca{} and \ha{} lines with the aim of maximising the correspondence between the two activity diagnostics and increase the \ha{} sensitivity to stellar signals at short and long timescales.
To our knowledge, this is the first time the impact of changing the bandwidth size of the \ha{} index is analysed for stars other than the Sun.
In \S \ref{sec:data} we present our sample and in \S \ref{sec:indices} we explain how we calculated the \ca{} and \ha{} indices.
In \S \ref{sec:correlations} we expose our methodology to determine the correlation coefficients and their significance.
We analyse the long-term correlations between \ca{} and \ha{} in \S \ref{sec:long_term} and the short-term correlations in \S \ref{sec:short_term}.
A discussion of the results in terms of stellar parameters, activity variability, and an analysis of the \ha{} line wings is presented in \S \ref{sec:discussion}, and we conclude in \S \ref{sec:conclusions}.

\section{Sample and HARPS data}\label{sec:data}
The sample was selected from the \citet{gomesdasilva2021} catalogue of chromospheric activity of 1\,674 FGK stars from the HARPS \citep{mayor2003} archive.
This catalogue used more than 180\,000 spectra to estimate precise and homogeneous mean activity levels and dispersion.
The catalogue also includes stellar atmospheric parameters such as effective temperature, metallicity and surface gravity along with isochronal masses, radii and ages.
The High Accuracy Radial Velocity Planet Searcher (HARPS) is a high-resolution, high-stability, fibre-fed, cross-dispersed echelle spectrograph with a resolution of $\lambda / \Delta \lambda = 115\,000$ and a spectral range from 380 nm to 690 nm, mounted at the ESO 3.6 m telescope in La Silla, Chile.
For a detailed description of the instrument we refer to \citet{pepe2002}.

\begin{figure}
	\resizebox{\hsize}{!}{\includegraphics{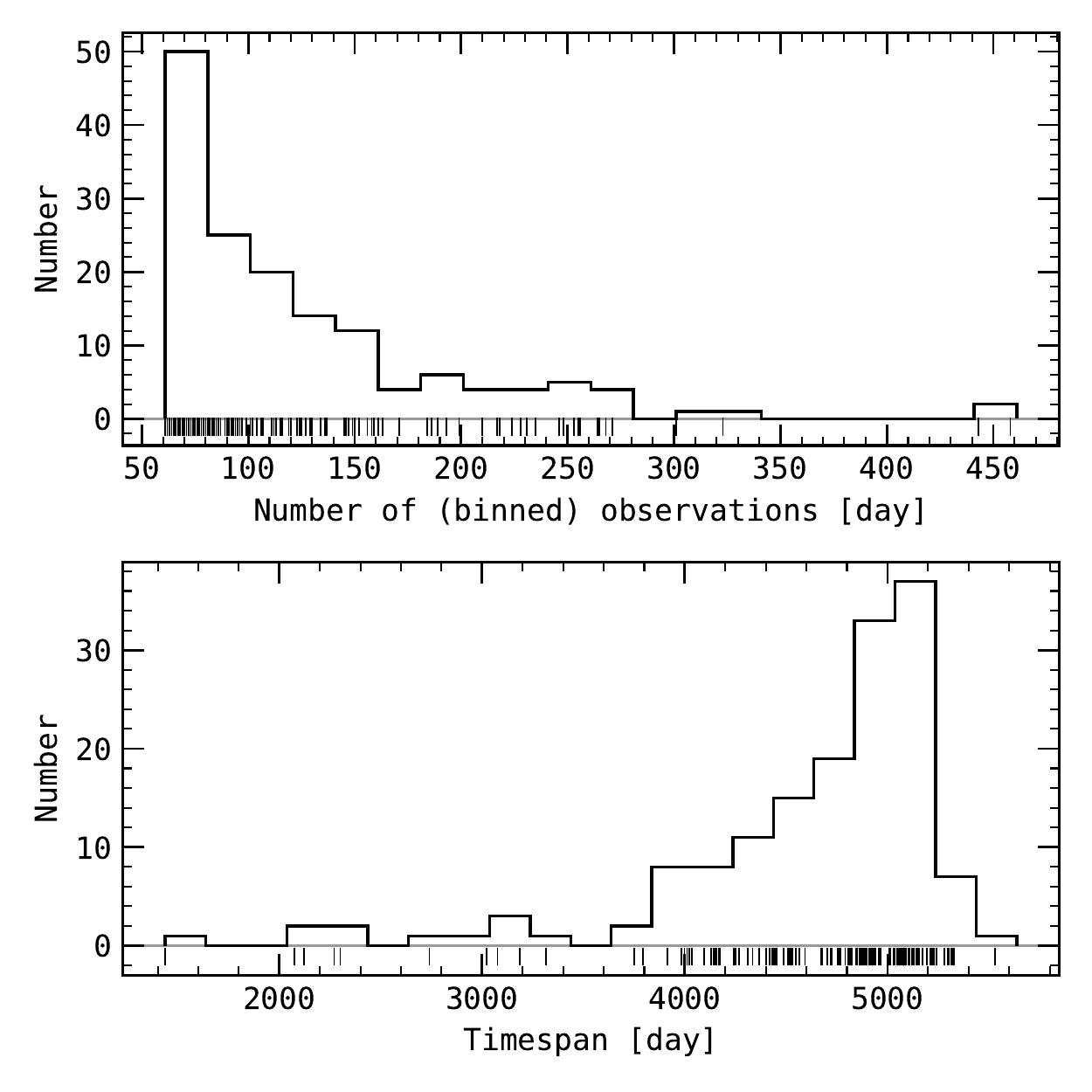}}

	\caption{
		Distribution of number of observations (upper panel) and timespan (lower panel) per star.
	}
	\label{fig:hist_sample_obs}
\end{figure}

The main objective of the present work involves the comparison of the activity signals measured in the flux of the \ca{} and \ha{} lines, and therefore our timespan and cadence needs to cover the rotational modulation and activity cycles related variability for each star, if possible.
To achieve this we need stars with high number of observations and long timespans.
We therefore selected from the catalogue main sequence FGK stars with more than 60 days of observations and timespans longer than 1\,000 days.
As we will see in \S\ref{sec:short_term} and can be observed in Fig. \ref{fig:ca_time_series}, this selection enabled us to identify activity minima and maxima in magnetic cycles for most stars, and to have enough data points to use those epochs to compare the behaviour of our activity indices.
Since some stars have very high cadences, characteristic of asteroseismology surveys, we imposed a maximum limit of 1\,000 spectra per star to reduce computational costs.
Outliers were removed via a sequential 4-sigma clipping of the \sca{}, \wsha{}, and \nsha{} indices time series (these indices are explained in \S \ref{sec:indices} and \S \ref{sec:correlations}).
Since we are only interested in timescales longer than one day, all data was daily binned to mitigate high frequency noise \citep{dumusque2011a} and reduce the datasets.
This selection resulted in a sample of 152 FGK dwarfs, including 101 G, 33 K and 18 F stars, with a total of 19\,019 binned data points.
Figure \ref{fig:hist_sample_obs} shows the distributions of the number of binned observations and timespan per star.
The number of binned observations per star range between 61 and 458 with the median at 101, and the timespan ranges from 1\,437 days (3.9 years) to 5\,531 days (15.1 years) with the median at 4\,865 days (13.3 years).

\begin{figure*}
	\resizebox{\hsize}{!}{\includegraphics{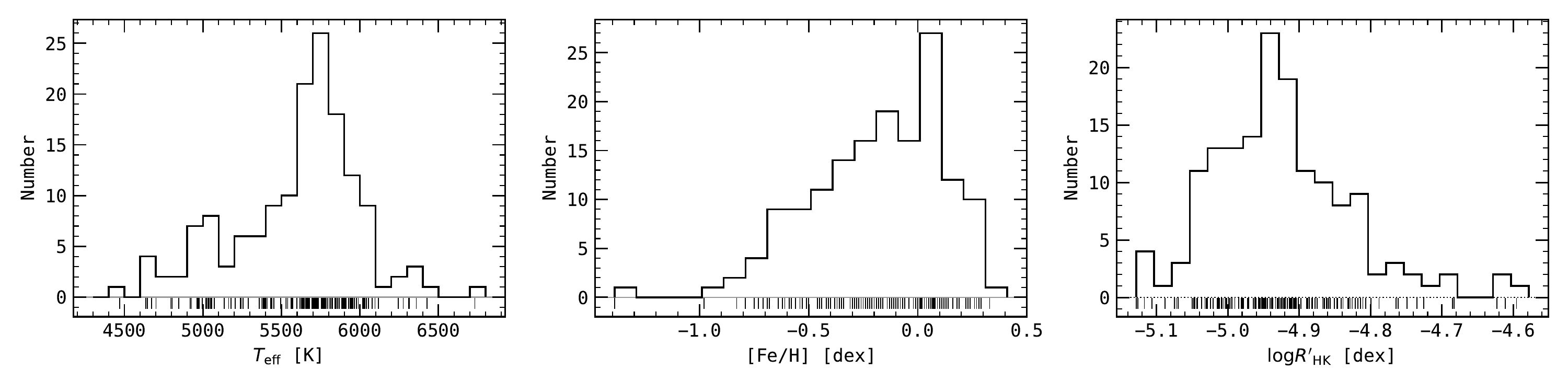}}

	\caption{
		Distribution of effective temperature (left panel), metallicity (middle panel), and \rhk{} activity level (right panel) for our sample.
	}
	\label{fig:hist_sample_params}
\end{figure*}

We also obtained stellar atmospheric parameters such as effective temperature, $T_\text{eff}$, metallicity, [Fe/H], and median chromospheric emission ratio, \rhk{}, from \citet{gomesdasilva2021} and references therein.
The distribution of these parameters for the sample are shown in Fig. \ref{fig:hist_sample_params} and their values provided in Table \ref{tab:stellar_table}.
The effective temperature varies between 4\,470 and 6\,732 K, with the median at 5\,694 K, the metallicity between $-1.39$ and 0.33 dex, with the median at $-0.11$ dex, and \rhk~ranges from $-5.13$ to $-4.60$ dex, with the median at $-4.93$ dex.
Thus, the majority of our sample stars have about 100 days of observation on a timespan of almost 5000 days ($\sim 13.7$ yr), and have effective temperature and activity levels similar to our Sun, with slightly sub-solar metallicity.
These biases are due to the majority of these stars coming from exoplanet surveys targeting inactive, solar-type stars.
Nevertheless, we should be able to assess the influence of these stellar parameters on the behaviour of the activity indices, if they are present.

\section{Activity indices based on the Ca II H\&K and H$\alpha$ lines}\label{sec:indices}
We used \verb+ACTIN+\footnote{\url{https://github.com/gomesdasilva/ACTIN}} \citep{gomesdasilva2018, gomesdasilva2021} to calculate the \ca~and \ha~indices for the full sample.
\verb+ACTIN+ calculates indices by integrating the flux in the activity sensitive lines and dividing them by the flux in pseudo-continuum regions:
\begin{eqnarray}
	I = \frac{\sum^{N}_{i}F_i}{\sum^{M}_{j}R_j}
\end{eqnarray}
where $F_i$ is the flux in the activity line $i$, $N$ is the number of activity lines (e.g. for \ca{} $N = 2$, for \ha{} $N = 1$), $R_j$ is the flux in the pseudo-continuum region $j$, and $M$ the number of pseudo-continuum regions. Generally there two pseudo-continuum regions ($M = 2$) surrounding the activity lines in the redder and bluer nearby continuum.
\verb+ACTIN+ uses linear interpolation to deal with the finite wavelength resolution of the spectrograph when integrating the flux over a given bandpass.
For more information related to the flux determination and errors we refer to \citet[][Appendix A]{gomesdasilva2021}.

The \ca{} index, which comes preinstalled in \verb+ACTIN+, was measured following the procedure in \citet{gomesdasilva2021} which mimics the original Mt.Wilson $S$-index derivation by \citet{vaughan1978}.
In the case of \ha{}, we used the reference lines from \citet{gomesdasilva2011} but computed the index for a range of central bandwidths between 0.1 and 2.0 \AA{}, in steps of 0.1 \AA{}.
Since we are not comparing the values of these indices between different stars, only the correlations between them, the temperature dependent photospheric contribution and bolometric correction are not required.
To compare the activity levels of stars we use the usual \rhk{} indicator \citep{noyes1984,rutten1984}.
From now on we will refer to the spectral lines as \ca{} and \ha{}, and to the indices as \sca{} and \sha{}.

\section{Correlation coefficients and significance}\label{sec:correlations}
As a metric for correlation we choose the Spearman coefficient over the Pearson so that we can infer correlations that are monotonic, either linear or not.
Furthermore, the Spearman coefficient does not require the two datasets to be normally distributed, as is the case where we have correlated time series with signals such as the quasi-periodic modulation due to activity.
The Spearman correlation coefficient, $\rho$, was calculated for \sca{} and each \sha{} using a different bandwidth.
To quantify the correlation significance, we calculated the $p$-value of the coefficients following the methodology described in \citet{figueira2013}\footnote{A python implementation of the algorithm is available at \url{https://bitbucket.org/pedrofigueira/line-profile-indicators/src/master/}.}.
Briefly, for each star, we performed 10\,000 Fisher-Yates shuffling of the data pairs (\sca{} and \sha{}) to create unbiased and uncorrelated datasets, for which we calculated the correlation coefficients.
The original correlation coefficient (before shuffling) was then compared with the mean of the shuffled population correlation coefficients, and the original dataset $z$-score was calculated.
The $p$-value, the probability of having an equal or larger correlation coefficient under the null hypothesis that the data pairs are uncorrelated, was obtained from the one-sided probability of having such a $z$-score from the observed Gaussian distribution.

To compare values of the coefficients for narrow and wide bands we use two representative bandwidths. 
As referred to in the introduction, the most used \ha{} bandwith to calculate the \ha{} index is the 1.6 \AA{} bandpass.
Due to its widespread use, we will consider it as an example of a wide bandpass, and will refer to this index as \wsha{} from now on.
As a reference to a narrow bandwidth we will use the \ha{} index calculated by integrating a 0.6 \AA{} bandwidth, referred to as \nsha{}\footnote{As we will see in \S \ref{sec:bw_best} this is the bandwidth that maximises the correlation between \sca{} and \sha{}.}.
For computational cost reasons, we only calculated the $p$-values for correlations using \nsha{}, \wsha{}, and $S_{\text{H}\alpha\text{W}}$ (this last index is discussed in \S \ref{sec:hawings}) both for the long- and short-term datasets.

From now on, we refer to "strong" correlations if $\rho$ has values higher than 0.5 in absolute value, and "weak or no correlation" if $\rho$ has values between $-0.5$ and $0.5$.
For example, we can have strong correlations that are insignificant ($p$-value $\geq 10^{-3}$) or weak correlations that are significant ($p$-value $\leq 10^{-3}$).
As we will see in the following sections, for the long-term datasets all the strong correlations are significant, however that is not always true for the short-term datasets.

\section{Long-term correlations} \label{sec:long_term}
To investigate the correlation between \sca{} and \sha{} in at long timescales we used the full time series for all stars.
The Spearman correlation coefficients and their $p$-values when using band-pass widths of 0.6 and 1.6 \AA{} as well as using the \ha{} line wings between 1.6 and 0.6 \AA{} (see \S \ref{sec:hawings}) are provided in Table \ref{tab:corr_table_long}.

\subsection{Correlations for different \ha{} bandwidths}
We expect the correlation between \ca{} and \ha{} to vary with the \ha~bandwidth used because different depths of the \ha~line probe different heights of the stellar atmosphere.

\begin{figure}
	\resizebox{\hsize}{!}{\includegraphics{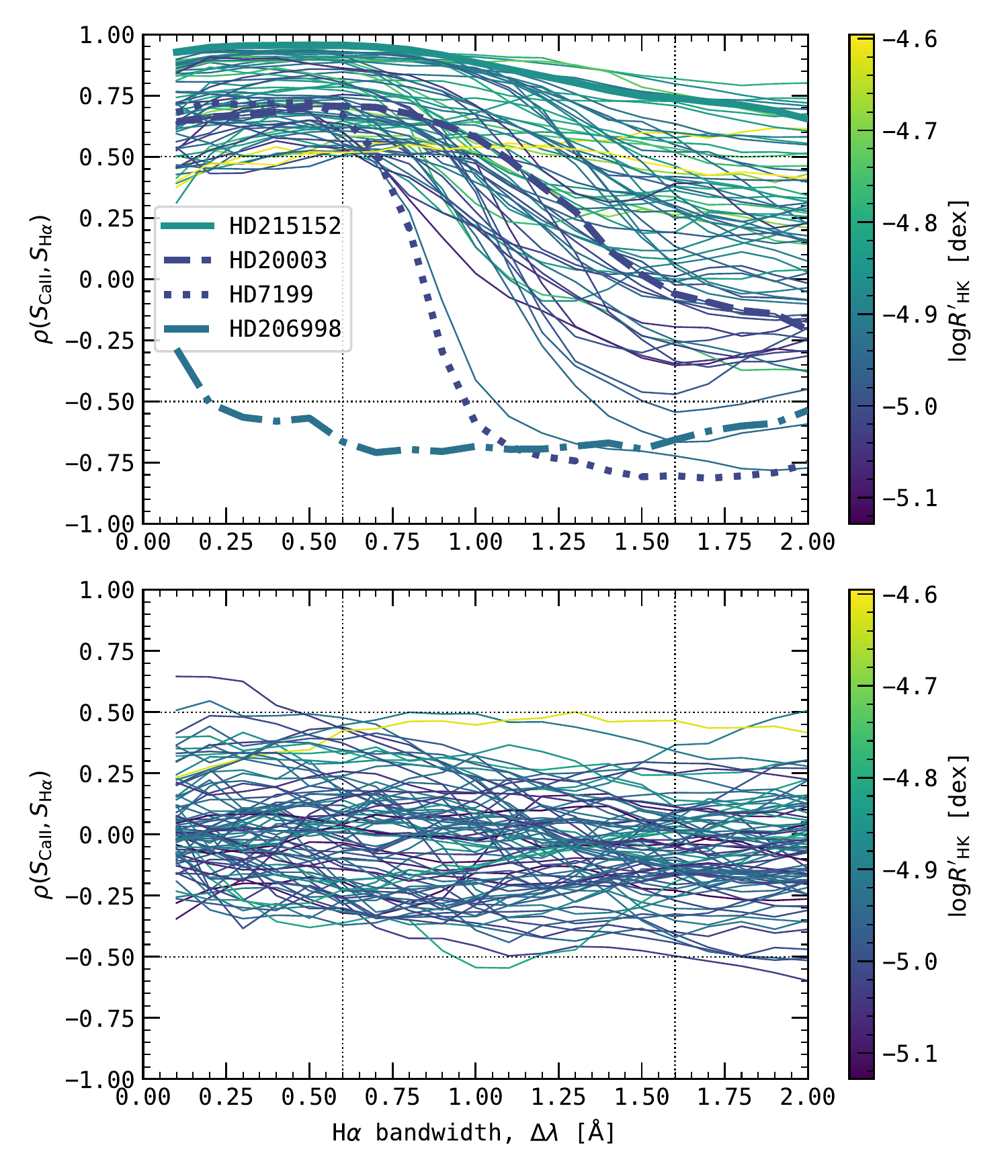}}

	\caption{
		Effect of varying the \ha{} bandwidth in the correlation between \ca{} and \ha{} coloured after activity level measured as \rhk.
		\textit{Upper panel:} Stars with strong positive ($\rho \geq 0.5$) or negative ($\rho \leq -0.5$) correlation with the \nsha{} index.
		Stars marked with thick lines are examples used in Figs. \ref{fig:ha_lines}, \ref{fig:example_ca_ha16_ha06}, and \ref{fig:example_ca_ha16_hawings}, and the only case of negative correlation with \nsha{} (HD\,206998).
		\textit{Lower panel:} Stars with weak or no correlation ($-0.5 < \rho < 0.5 $) with the \nsha{} index.
		In both panels, as an indication, we mark the boundaries between strong and no correlation with horizontal dotted lines at $\rho$ values of 0.5 and $-0.5$ with vertical dotted lines.
	}
	\label{fig:rho_bw_examples}
\end{figure}

Figure \ref{fig:rho_bw_examples} shows the Spearman correlation coefficients as a function of \ha{} bandwidth  coloured after activity level.
In the upper panel we show only stars having strong positive or negative correlations ($\rho \geq 0.5$ and $\rho \leq -0.5$) using the 0.6 \AA{} bandwidth as an example of a narrow bandpass, while in the lower panel only stars with weak or no correlations ($-0.5 < \rho < 0.5$) with \nsha{} are represented.
In the case of \nsha{} and \wsha{}, all stars with $|\rho| \geq 0.5$ also have $p$-values $\leq 10^{-3}$, meaning that all those correlations are statistically significant.
On a first observation, we see a tendency for the correlation to decrease as the \ha{} bandwidth increases.
Four different behaviours of the correlation as a function of bandwidth can be observed: (1) "flat-positive", where the correlation is strong positive and almost constant across bandwidths, (2) "positive-zero", where it decreases from strong positive to weak or no correlation, (3) "positive-negative", where it decreases from strong positive to strong negative, and (4) "flat-negative", where it is constant but always strong negative.
The figure also provides other interesting information:
\begin{itemize}
	\item All cases of strong positive \sca{}--\wsha{} correlation also have strong positive \sca{}--\nsha{} correlation (upper panel) -- there are no cases where the correlation increases from weak to strong with increasing bandpass width.
 	\item All cases of strong negative \sca{}--\wsha{} correlation have strong positive \sca{}--\nsha{} correlation, except for HD\,206998 (upper panel) -- narrow bandpasses can turn strong anti-correlations into strong positive correlations.
   	\item All cases of weak or no correlation between \sca{} and \nsha{} also show weak or no correlation between \sca{} and \wsha{} (lower panel) -- these stars never show strong correlations between the calcium and hydrogen lines regardless of the \ha~bandwidth used.
   	\item The only case of strong negative \sca{}--\nsha{} correlation, HD\,206998, also has strong negative \sca{}--\wsha{} correlation (upper panel) -- since we only have one star in this situation we can not ascertain whether this is an outlier or an additional "behaviour".
   	\item In general, stars with lower correlations with \wsha{} tend to have lower activity levels (upper panel) -- in agreement with \citet{gomesdasilva2014, meunier2022}.
\end{itemize}
These behaviours show that, for most inactive\footnote{Active stars are known to have strong positive correlations between \sca{} and \sha{}} FGK dwarfs, using a narrow \ha{} bandpass can improve the correlation between the \ha{} and \ca{} indices, and that most of the anti-correlations detected between these two indices are caused by using a wide bandwidth when calculating the \ha{} index.
It is also interesting to note that if a star has no correlation using \nsha{}, then widening the bandpass will not improve the correlation into either strong positive or negative values.
In general terms, narrower bandwidths are preferred if one wants to detect activity signals similar to those followed by \ca{} (which is known to correlate well with plages in the Sun and with activity induced radial velocity).

\begin{figure*}
	\resizebox{\hsize}{!}{\includegraphics{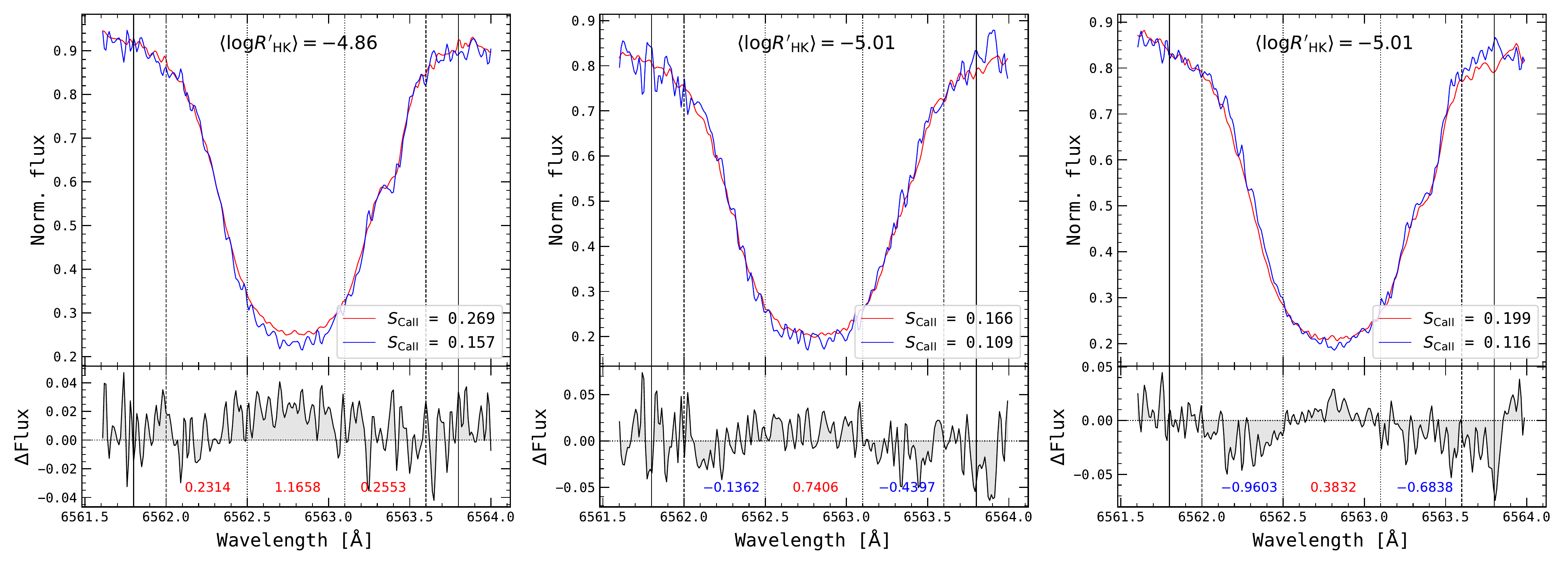}}

	\caption{
		\textit{Upper panels:} \ha{} line profiles of HD\,215152 (left), HD\,20003 (middle), and HD\,7199 (right) at their maxima (red) and minima (blue) activity levels.
		\textit{Lower panels:} Difference between the fluxes at maximum and minimum of activity for each star.
		Vertical lines show bandwidths of 2.0 \AA{} (solid), 1.6 \AA{} (dashed), and 0.6 \AA{} (dotted). The numbers in the lower panels show the integrated flux difference in each region delimited by the vertical lines in red if positive and blue if negative.
	}
	\label{fig:ha_lines}
\end{figure*}

In Fig. \ref{fig:ha_lines} we show three examples of the \ha{} profile for three cases represented in Fig. \ref{fig:rho_bw_examples} (upper panel) in which all have similar \sca{}--\nsha{} correlation coefficients but behave differently as the \ha~bandwidth is increased.
To compare the \ha{} lines at maxima and minima we normalised the spectra by the mean of the flux in the $R_1$ and $R_2$ pseudo-continuum regions used to compute the \sha{} index.
The left panel shows HD\,215152 (K3V), a star with a "flat-positive" behaviour, in the middle panel we show HD\,20003 (G8V), a star showing "positive-zero" behaviour, and in the right panel HD\,7199 (G9V) an example of a "positive-negative" correlation behaviour.
The top panels show the \ha{} line at the minimum (blue) and maximum (red) of \sca{} activity for each star while the bottom panels show the difference in flux between these extremes.
The vertical lines in all figures represent the limits of the 0.6 \AA{} bandwidth (dotted lines), the 1.6 \AA{} (dashed lines), and the maximum bandwidth value we used of 2.0 \AA{} (solid lines).
The numbers in the lower panels show the flux difference calculated in each width, coloured red if positive (more flux at activity maximum) and blue if negative (more flux at activity minimum).
The first thing we can observe in these examples is that the flux in the core always increases with \sca{} activity level, while the flux in the wings can increase or decrease with activity level.
We can see that for the case of HD\,215152 (left panel), as \ca{} increases, both the \ha{} line core (between the 0.6 \AA{} limits) and the lateral "wings" of the lines increase in flux.
Thus, the correlation between \ca{} and \ha{} is positive, either if using a narrow or a wider bandpass.
On the other hand, in the case of HD\,20003 (middle panel), while the difference in flux in the line core is positive as the activity increases, the line wings widen and thus decrease in flux, producing a negative flux difference.
This means that, while we get a positive correlation using a narrowband, as we increase the bandwidth, we start degrading the correlation by including the negative flux difference from the line wings, and the correlation drops to almost zero because the positive flux difference in the core and the negative flux difference in the wings almost compensate each other.
If the broadening (the drop in flux) of the \ha{} wings is larger than the increase in flux in the core, like in the case of HD\,7199 (right panel), the correlation can become negative when a wide bandpass is used.
However, a strong positive correlation is still possible if a narrow bandwidth is used instead.
A similar \ha{} profile behaviour was previously observed by \citet{flores2016, flores2018} while studying the activity cycles of the two solar analogs HD\,45184 and HD\,38858 (both included in our sample). While the \ha{} equivalent width showed no significant variability, the shape of the \ha{} profile produced variations compatible with the \ca{} cyclic variations. In fact, their \ha{} profile core (<0.7 \AA{} from centre) increased in flux with \ca{} activity level while the "wings" flux decreased.

\subsection{\ha{} bandwidth that maximises correlations}\label{sec:bw_best}
Now that we have a strong indication that narrower bandwidths tend to improve the \ca{}-\ha{} correlation, we are going to determine which bandpass maximises it.

\begin{figure}
	\resizebox{\hsize}{!}{\includegraphics{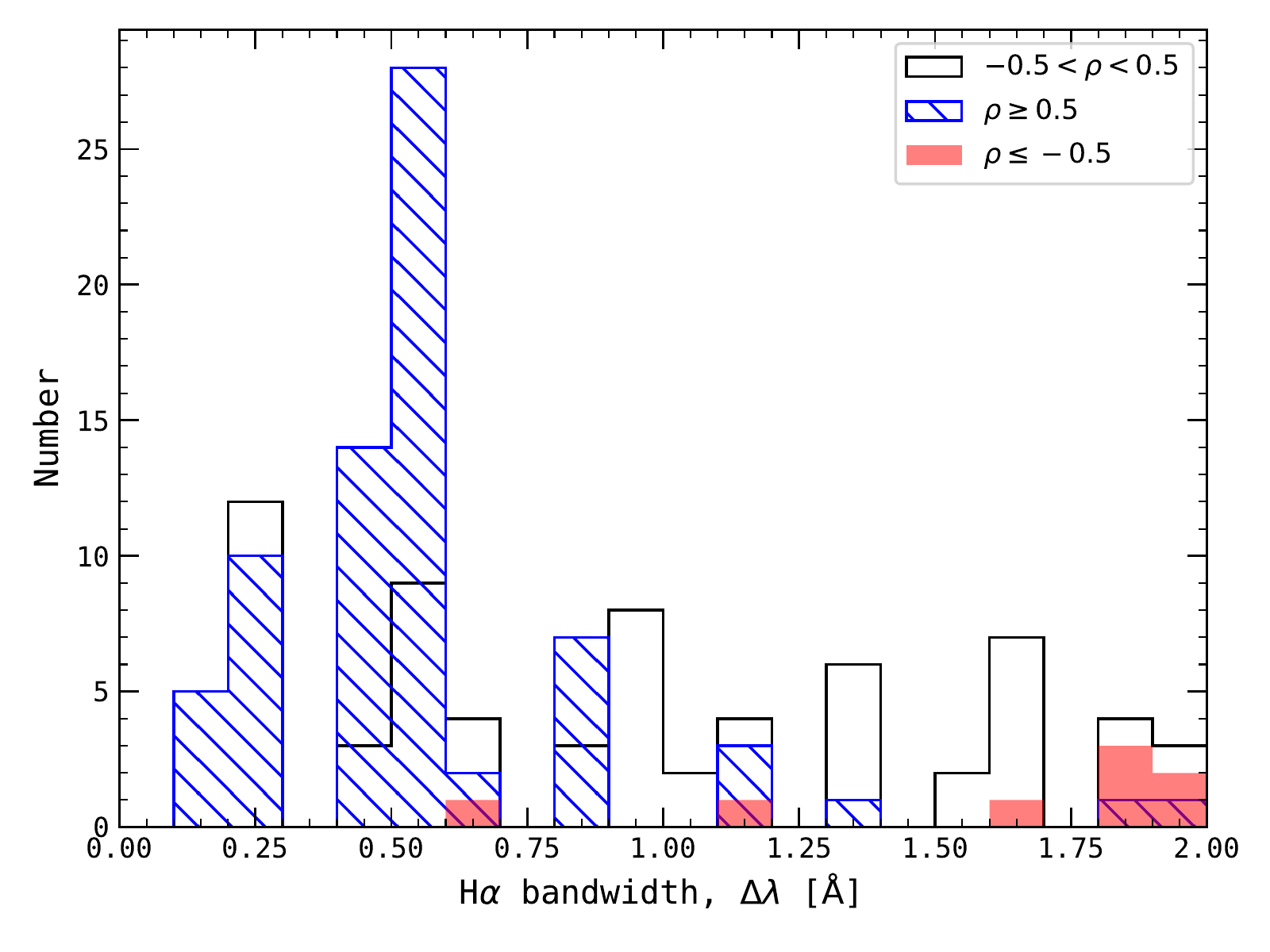}}

	\caption{
		Distribution of bandpasses that maximise the correlation between \ha{} and \ca{}.
		The black histogram shows the bandpass distribution for stars with weak or no correlation coefficient ($-0.5 < \rho < 0.5$), the blue hatched histogram shows stars with strong positive correlations ($\rho \geq 0.5$) and the red histogram stars with strong negative correlations ($\rho \leq -0.5$).
	}
	\label{fig:bp_best}
\end{figure}

For each star, we selected the bandwidth for which the correlation coefficient, $\rho$, has maximum \textit{absolute} value, thus maximising either positive or negative correlations.
Figure \ref{fig:bp_best} shows the distribution of bandpasses that maximise the correlation, where the black histogram shows the bandpass distribution for stars for which the maximum absolute $\rho$ never passes the strong correlation threeshold, the blue hatched and red histograms shows the bandpass distribution for stars for which the maximum $\rho$ is either strong positive or negative, respectively.
For the 152 stars in this sample we found an equal number of stars with strong positive and weak or no correlations, 72 (47.4\%), while only 8 stars (5.3\%) show strong negative correlations.
The black histogram shows that there is a weak tendency for the bandpasses that maximise the correlation between the two indices to have narrower widths even though the correlation is never strong.
When we select only the maximum strong correlations, either positive (blue hatched) or negative (red) it becomes clear that using narrower filters between around 0.4 and 0.6 \AA{} maximises the cases of strong positive correlations between \ha{} and \ca{}, while the few cases of strong anti-correlations (negative) appear preferably when using wider $>1.0$ \AA{} \ha{} bandpasses.
This maximisation process tends to prefer narrow bandpasses when the \ha{} wings flux behaves in such a way that it will compensate the flux changes in the core, and therefore including the wings will degrade the correlation coefficient absolute value, while it prefers wider bandpasses when the wings flux variation is larger than that of the core, thus having more importance when the line flux is fully integrated.

\begin{figure}
	\resizebox{\hsize}{!}{\includegraphics{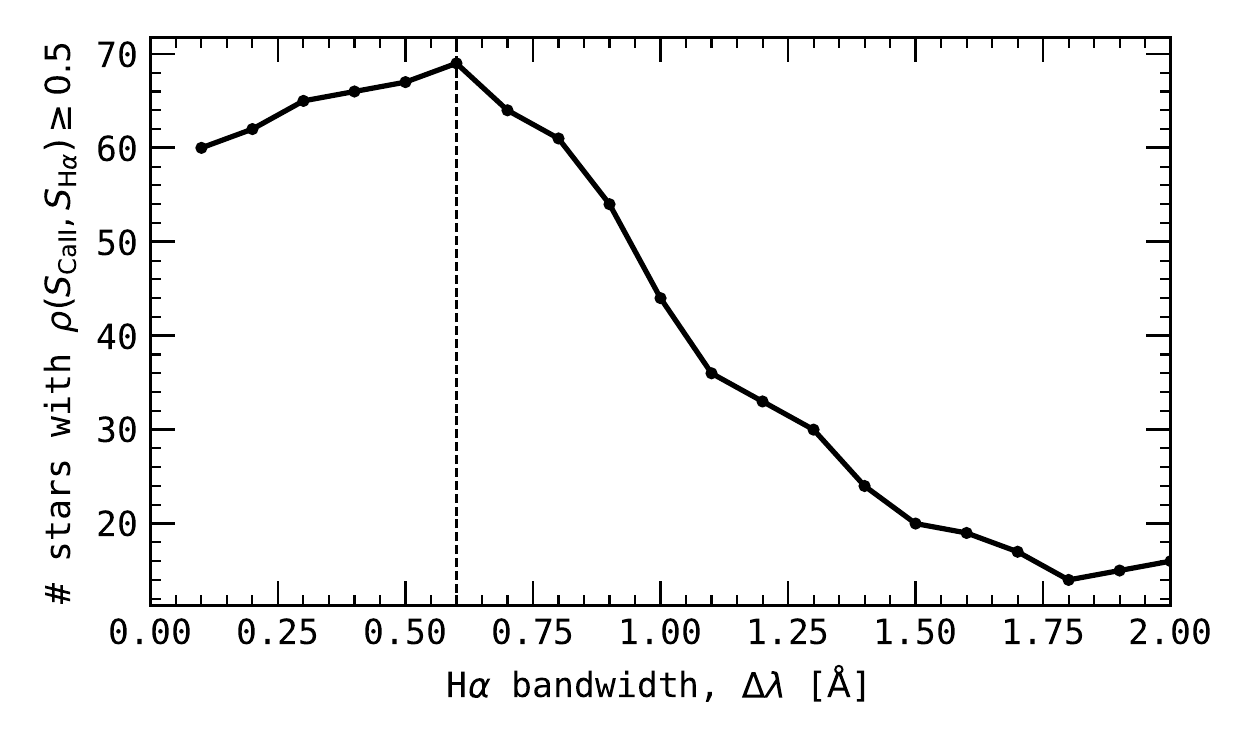}}

	\caption{
		Number of stars with correlation coefficient between \ca~and \ha{} greater than 0.5 (strong positive correlation) for different bandwidths.
		The vertical dashed line indicates the maximum at $\Delta \lambda = 0.6$ \AA{}.
	}
	\label{fig:n_stars_pos_bw}
\end{figure}

To identify the bandwidth that maximises the strong positive \ca{}-\ha{} correlation we calculated the number of stars with a correlation coefficient higher than 0.5 for each \ha{} bandwidth (Fig. \ref{fig:n_stars_pos_bw}).
The best bandwidth, $\Delta \lambda = 0.6$ \AA{}, was selected as the one including more stars.
The number of stars with strong positive correlation using 0.6 \AA{} is 69 (45\%) while using the widely used 1.6 \AA{} bandpass is just 20 (13\%).
Even though the optimal bandwidth is 0.6 \AA{}, any bandwidth between around 0.25 and 0.75 \AA{} have similar results. Beyond $\sim$0.75 \AA{}, the number of stars with strong positive correlations starts to decline significantly, and those bandwidths should not be used to integrate the \ha{} flux for the \sha{} index.

\begin{figure*}
	\resizebox{\hsize}{!}{\includegraphics{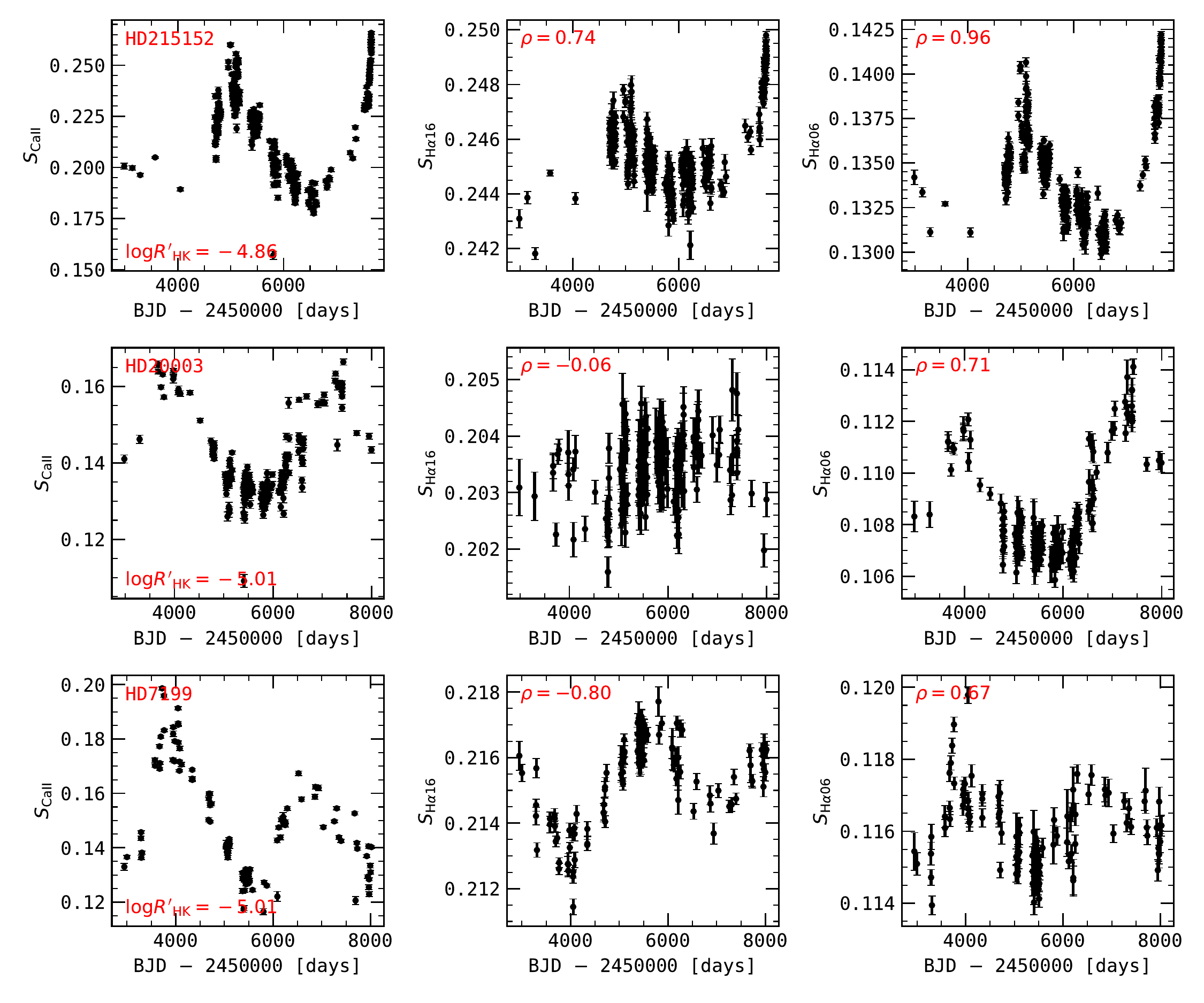}}

	\caption{\sca{}, \wsha{}, and \nsha{} for HD\,215152 (top panels), HD\,20003 (middle panels), and HD\,7199 (lower panels).}

	\label{fig:example_ca_ha16_ha06}
\end{figure*}

As an example of the benefits of using a 0.6 \AA{} bandwidth instead of 1.6 \AA{} to follow long-term activity we show in Fig. \ref{fig:example_ca_ha16_ha06} the time series of the three stars used as example in Figs. \ref{fig:rho_bw_examples} and \ref{fig:ha_lines}, namely HD\,215152, HD\,20003, and HD\,7199, using \sca{}, \wsha{}, and \nsha{} which show strong positive, no correlation, and strong negative correlations between \sca{} and \wsha{}, respectively.
This shows that using \nsha{} can increase the positive correlation observed in \wsha{} (top panels), turn no correlations (and an apparent flat signal) into a strong positive correlation with an obvious cycle pattern (middle panel), and turn a strong negative correlation (an "anti-cycle") into a strong positive correlation (a cycle, lower panel).
Two of these stars (HD\,215152 and HD\,7199) were used in \citet[][Fig. 5]{meunier2022} in their study of the correlation between the \sca{} and \sha{} indices as examples of positive and negative correlations for their \sha{} using 1.6 \AA{}.

\section{Short-term correlations}\label{sec:short_term}
After arriving to the conclusion that the bandwidth of $0.6$ \AA{} maximises the correlation between \ca{} and \ha{} for long timescales, we are now interested in assessing what happens in the short-term, more characteristic of the timescales of stellar rotation.
Furthermore, we are interested to see how is the behaviour of the \ha{} index, using wide or narrow bandwidths, when the stars are at the maximum and minimum of their activity cycles.
To achieve this, we need to identify the epochs of activity minima and maxima in the time series and then calculate the correlation coefficient between the \sca{} and \nsha{} and \ca{} and \wsha{} indices for these epochs.
Main sequence FGK stars have rotation periods that range from <1 day to around 60-70 days, depending on their ages \citep{mcquillan2014}, while the magnetic cycles of these types of stars are of the orders of years to decades \citep{baliunas1995}. We therefore need to select the maxima and minima epochs with timespans long enough to be able to cover the range of rotation periods of these stars while being short enough to isolate the maxima of activity from the minima and to exclude the effect of long-term variations associated with magnetic cycles.
The maxima and minima epochs were identified using the following methodology:
\begin{enumerate}
	\item Grouped the \sca{}, \nsha{}, and \wsha{} time series into a grid of 30 day steps. This value, close to the higher envelope of the rotation periods of FGK dwarfs \citep[][Fig. 1]{mcquillan2014}, is enough to cover most of the rotation periods in just one grid step.
	\item Each epoch was defined as a group of points surrounded by empty steps, so that, for example, two (30 day) consecutive steps surrounded by two empty (30 day) steps would constitute an epoch (with a 60 days step).
	\item Only epochs with at least seven observations were considered to ensure we have minimum points to calculate correlation coefficients.
	\item The epoch with the highest \sca{} mean value is the epoch at the activity maximum, while the epoch with the lowest \sca{} mean is the activity minimum.
	\item To ensure that the activity levels of the maximum and minimum epochs are well separated in activity level, we imposed that the difference between the means of each epoch is at least two times the average standard deviations of the two epochs: $\langle S_{\text{CaII}_\text{max}} \rangle - \langle S_{\text{CaII}_\text{min}} \rangle \geq 2 \langle \sigma( S_{\text{CaII}_\text{max}}), \sigma(S_{\text{CaII}_\text{min}}) \rangle$.
\end{enumerate}
This resulted in 103 stars with well separated epochs of activity maxima and minima.
The timespans obtained for the maxima and minima epochs range between 6 and 238 days, with the median at 88 days.
The \sca{} time series for these stars, with the maxima and minima identified with red and blue points, respectively, are provided in the Appendix \ref{app:plots_ca_time_series}.
After completing this selection process we calculated the Spearman correlation coefficients between \sca{} and \nsha{}, and \wsha{}, and their respective $p$-values, for the maximum and minimum epochs of each star.
The correlation coefficients along with $p$-values and mean and standard deviation of \sca{} for the full sample (before applying point 5 of the above selection) are provided in Table \ref{tab:corr_table_short}.
Contrary to the case of the long-term correlations in \S \ref{sec:long_term}, not all stars with $|\rho| \geq 0.5$ have $p$-values $\leq 10^{-3}$, which means that there are cases of strong correlations that are not statistically significant.
This is probably a consequence of the reduced number of data points in these datasets when compared to the full datasets used in the long-term analysis.

\begin{figure}
	\resizebox{\hsize}{!}{\includegraphics{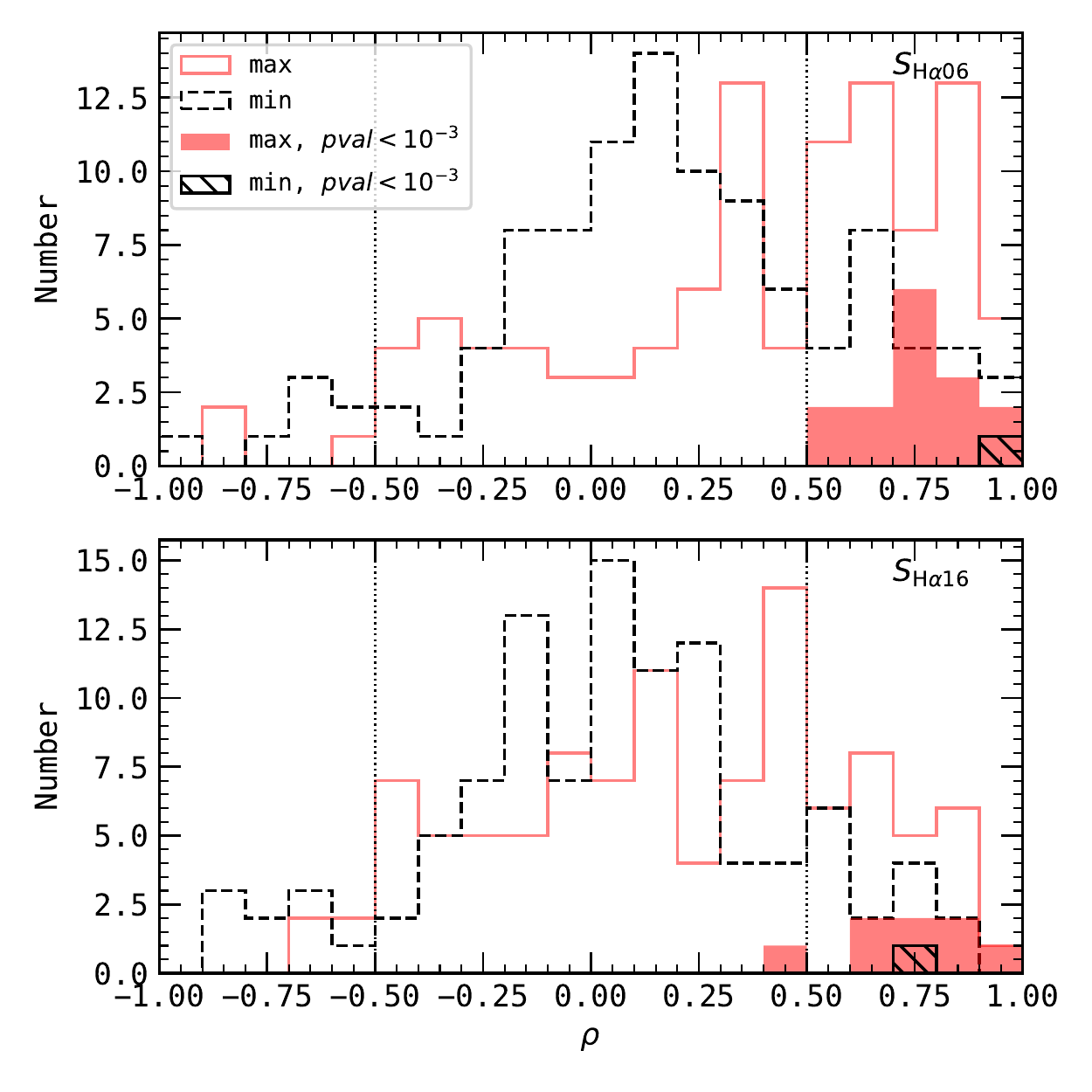}}

	\caption{
		Distribution of the correlation coefficient between \sca{} and \nsha{} (upper panel) and \sca{} and \wsha{} (lower panel). The red histograms are the coefficients for epochs at the maximum of activity ($p$-value $\leq 10^{-3}$, filled red) while the black histograms are the coefficients for epochs at minimum ($p$-value $\leq 10^{-3}$, hatched black).
	}
	\label{fig:hist_max_min}
\end{figure}

Figure \ref{fig:hist_max_min} presents the distribution of the correlation coefficients between \sca{} and \sha{} at maxima (red line) and minima (black dashed line) for the cases of \nsha{} (top panel) and \wsha{} (bottom panel).
The correlations with significant coefficients ($p\text{-value} \leq 10^{-3}$) are represented by the red filled histograms for the activity maxima and black hatched histograms for the activity minima.
The first thing to note is that the only cases of significant correlations, for both \nsha{} and \wsha{}, are positive correlations with $\rho$ close to or above 0.5.
There are more cases of strong positive correlations ($\rho \geq 0.5$) at maximum than at minimum of activity for both \ha{} indices, while there are more cases of strong negative correlations ($\rho \leq -0.5$) at minimum than at maximum.
Similarly to the long-term case, using the $0.6$ \AA{} bandwidth in \ha{} maximises the number of stars with strong positive correlation with the calcium lines, both for epochs at maximum and minimum of activity, however the number of significant ($p$-values $\leq 10^{-3}$) at minimum (black hatched histogram) is the same for the two indices.

Out of the 103 stars analysed, 50 (49\%) show strong positive correlations between \sca{} and \nsha{} at activity maximum (15, 15\%, having $p$-value $\leq 10^{-3}$), while 23 (22\%) have strong positive correlation at activity minimum (1, 1\%, having $p\text{-value} \leq 10^{-3}$).
There are no cases of significant negative correlations between \sca{} and \nsha{} at maximum or minimum of activity.
In comparison, when using \wsha{}, only 26 (25\%) showed strong positive correlations (7, 7\%, with $\text{p-value} \leq 10^{-3}$) at activity maximum and 14 (14\%) at activity minimum (1, 1\%, with $\text{p-value} \leq 10^{-3}$).
The figure also shows that at these shorter timescales there are more cases of anti-correlations both at minimum and maximum of activity for \nsha{}.
We should note, however, that we are using considerably fewer data points to compute the correlations at short timescales, and this will influence the significance of the correlation coefficients as demonstrated by the reduced number of cases with $p$-values below $10^{-3}$.

\begin{figure}
	\resizebox{\hsize}{!}{\includegraphics{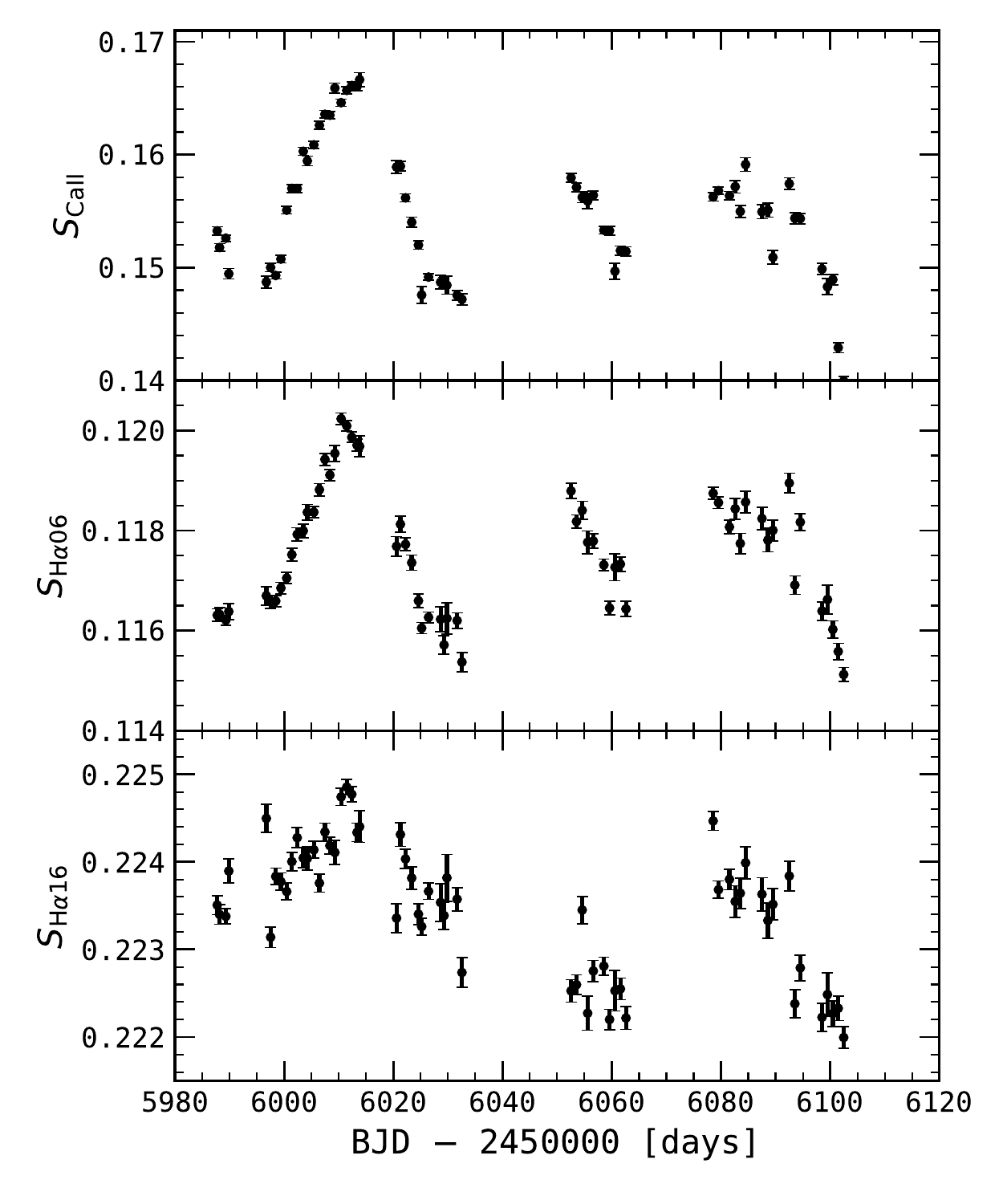}}

	\caption{
		Rotational modulation near the activity cycle maximum of HD\,109200 measured by the \sca{} (upper panel), \nsha{} (middle panel), and \wsha{} (lower panel) activity indices.
	}
	\label{fig:rot_mod}
\end{figure}

An example of the improvement of using the narrower bandwidth also at short timescales is given in Fig. \ref{fig:rot_mod} where we show the rotational modulation of the K dwarf HD\,109200 as measured by \sca{} (upper panel), \nsha{} (middle panel), and \wsha{} (lower panel), selected at a region close to the activity cycle maximum.
As can be observed, the \nsha{} signal closely follows the modulation observed in \sca{} while the signal from \wsha{} appears to degrade the modulation signal.
This has implications for the detection of rotation periods of stars when using the \ha{} time series. If a wide \ha{} bandwidth is used, the rotational modulation could be degraded, diminishing the significance of the signal. In the case of RV analysis in the context of exoplanet detection, this could increase the rate of false positives, by failing to identify activity signals, their harmonics and aliases.

\section{Discussion}\label{sec:discussion}

\subsection{Trends with effective temperature, metallicity and activity level}
Although using a 0.6 \AA{} bandwidth maximises the correlation between \sca{} and \sha{}, there are still about half of the stars which show weak or no correlation.
To understand why, we searched for clues by analysing the correlation coefficient between \sca{}, \nsha{} and \wsha{} against the stellar parameters effective temperature, metallicity and the activity level measured by \rhk.

\begin{figure}
	\resizebox{\hsize}{!}{\includegraphics{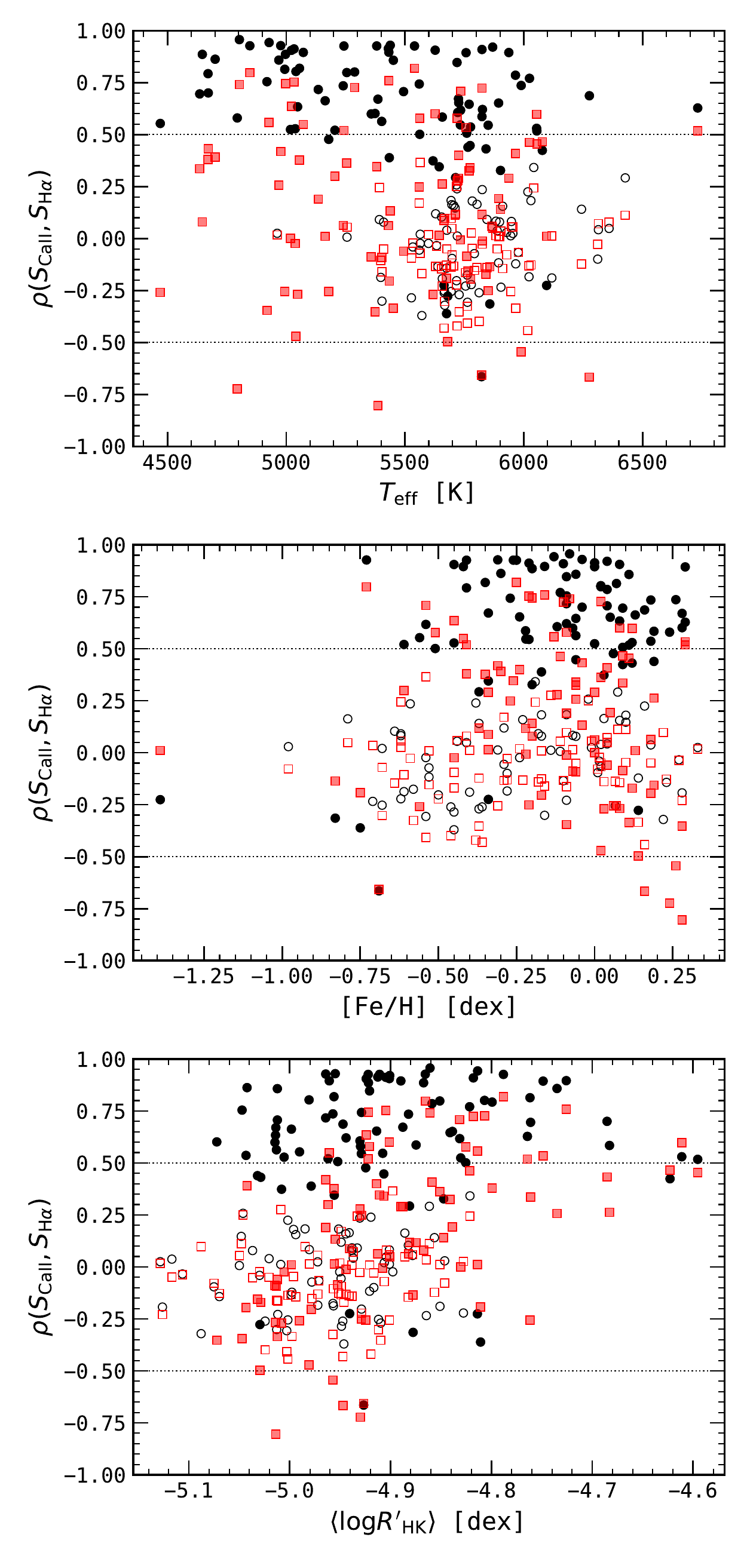}}

	\caption{Spearman correlation coefficient between \sca{} and \nsha{} (black circles) and \sca{} and \wsha{} (red squares) against effective temperature (top panel), metallicity (middle panel), and median \rhk{} activity level (bottom panel).
	Horizontal dotted lines mark the thresholds for strong correlations at $\rho = 0.5$ and $-0.5$.
	Filled markers are significant correlations with $p$-value $\leq 10^{-3}$.
	}
	\label{fig:rho06_params}
\end{figure}

The top panel of Fig. \ref{fig:rho06_params} shows that there is no trend of the \sca{}-\wsha{} correlation (red squares) with effective temperature.
This was also observed by \citet{gomesdasilva2014} and \citet{meunier2022} using a 1.6 \AA{} \ha{} bandpass.
However, when using \nsha{} as the \ha{} index (black circles) we see that almost all K dwarfs ($T_\mathrm{eff} < 5250$ K) have strong positive correlations, while hotter F and G stars have strong positive ($\rho \geq 0.5$), weak or no correlations ($-0.5 \leq \rho \leq 0.5$), and strong negative correlations ($\rho \leq -0.5$).

In the middle panel, when using \wsha{} (red squares) there is a tendency for the strong negative correlations ($\rho \leq -0.5$) to be present at higher metallicity values, as was previously observed by \citet{gomesdasilva2014}.
This trend disappears when using \nsha{} (black circles), meaning that the correlation between \sca{} and \nsha{} is not related to the metal content of stars.

The impact of the \ca{} activity level in the \sca{}-\wsha{} correlation was already observed for FGK dwarfs by \citet{gomesdasilva2014} and \citet{meunier2022} and for M dwarfs by \citet{walkowicz2009} and \citet{gomesdasilva2011}, where more active stars have a tendency to show strong positive correlations, while negative and weak or no correlations appear preferably for inactive stars.
The bottom panel shows this tendency is present when using \wsha{} (red squares).
However, for a narrow bandpass on \ha{} (black circles), we find that the strong positive correlations can be present for a greater number of inactive stars, including those with \rhk levels below $-5.0$ dex.
There is also no strong negative correlations when using \nsha{} (except one possible outlier).
The correlation between \sca{} and \nsha{} seems to be independent on the activity level of the stars.

\subsection{The effect of activity variability}

\begin{figure*}
	\resizebox{\hsize}{!}{\includegraphics{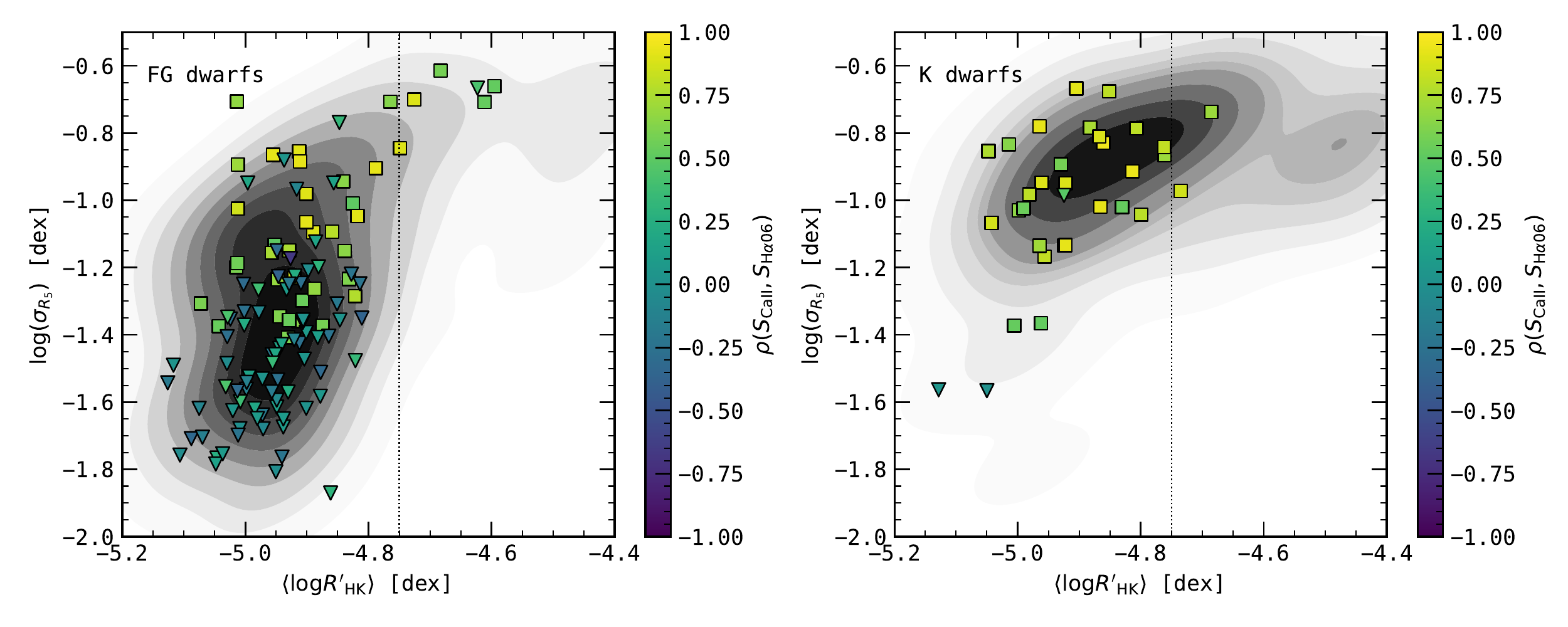}}

	\caption{
		Stellar activity variability measured as $\log (\sigma_{R_5})$ versus median activity level, $\langle$\rhk{}$\rangle$, for our sample coloured after the Spearman correlation coefficient between \sca{} and \nsha{}. 
		Left panel are FG dwarfs while the right panel shows K dwarfs.
		Squares are stars with $\rho \geq 0.5$ (strong positive correlations) while inverted triangles stars with $\rho < 0.5$.
		The gray scale map shows the Bivariate Kernel Density Estimate (KDE) from the catalogue of \citet{gomesdasilva2021} using just the main sequence stars of the selected spectral types of each panel.
		The vertical dotted line at \rhk{}$ = -4.75$ dex marks the original Vaughan-Preston gap \citep{vaughan1980}, separating the active and inactive stars.
	}
	\label{fig:kde_rho06}
\end{figure*}

\citet{gomesdasilva2021} studied the chromospheric activity of FGK stars and found that the vast majority of the K dwarfs in their sample (including stars and observations used here) have high \rhk{} variability even though they are considered inactive, with \rhk{} $< -4.75$ dex.
The authors also observed that F and G inactive dwarfs can have both low and high variability and that FGK dwarfs with activity levels above around $-4.8$ dex tend to have high variability.
We therefore suspect that the explanation for the lack of stars with weak or no correlation for higher activity levels (\rhk{} $> -4.8$ dex) and for K dwarfs ($T_\text{eff} < 5250$ K) might be the same: high activity variability.
To test this, we plotted the activity variability-level diagram with a 2D density map from the \citet{gomesdasilva2021} catalogue using just main sequence stars and overplotted the stars from this sample using squares for stars with strong positive correlation ($\rho \geq 0.5$) and inverted triangles for stars with weak or no, or negative correlation ($\rho < 0.5$), coloured after their $\rho$ value, as shown in Fig. \ref{fig:kde_rho06}.
As we can see, the majority of the stars with strong positive correlation have high activity variability (with $\log (\sigma_{R_5}) > -1.4$ dex) while the stars with weak or no correlation have low variability.
Furthermore, activity level is not a main factor for strong positive correlations between \sca{} and \nsha{}, since they are present in both inactive and active stars, as can also be observed in Fig. \ref{fig:rho06_params}, lower panel.

Figure \ref{fig:kde_rho06} also shows why we found previously that almost all K dwarfs have strong positive correlations (right panel) while some FG dwarfs show weak or no correlations (left panel): almost all K dwarfs have high activity variability while F and G dwarfs show a greater range in variability that reach lower amplitudes.
This shows that, instead of activity level, the \ca{} variability (here measured by the logarithm of the dispersion of $R_5 = 10^5 \times R'_\text{HK}$) is one of the main contributions for the correlation between \ca{} and \nsha{}.

\subsection{The \ha{} wings}\label{sec:hawings}
In the previous sections we showed that the inclusion of flux from the \ha{} wings in the \sha{} index increases the number of strong negative correlations between \sca{} and \sha{}, and deteriorates the activity signals, especially for stars with lower activity levels and higher metal content.
To inspect the behaviour of the \ha{} wings with \sca{} activity, we constructed a new index, $S_{\text{H}\alpha\text{W}}$, using the flux of \ha{} considering a bandpass with limits between those of 1.6 and 0.6 \AA{}.
The Spearman correlation coefficient between \sca{} and $S_{\text{H}\alpha\text{W}}$ and the corresponding $p$-values were calculated according to the methodology described in \S \ref{sec:correlations}.

Most of the significant correlations between $S_{\text{H}\alpha\text{W}}$ and \sca{} are negative and these are present mainly for inactive stars with \rhk below around $-4.9$ dex and stars with higher metallicity (Fig. \ref{fig:rho06_params_wings}).
This behaviour explains the anti-correlations between \sca{} and \wsha{} in inactive stars \citep{gomesdasilva2014, meunier2022} and also the tendency for these to be present in higher metallicity stars \citep{gomesdasilva2014}.
Interestingly, there are also a few stars with positive correlations between \sca{} and $S_{\text{H}\alpha\text{W}}$, with $\rho$ close to 0.5.
These stars are typically G dwarfs with higher activity levels than the ones with negative correlations.

Figure \ref{fig:example_ca_ha16_hawings} shows the comparison between the time series of \sca{}, \wsha{} and $S_{\text{H}\alpha\text{W}}$ for the same stars used before to exemplify the three different behaviours of the \ha{} index when compared to \sca{}.
For HD\,215152 (upper panels), we saw in Fig. \ref{fig:example_ca_ha16_ha06} that both the \nsha{} and \wsha{} indicators were positively correlated with \sca{}, with \nsha{} having a stronger correlation coefficient than \wsha{}.
Here we see that the index based solely on the \ha{} wings flux has a very weak correlation with \sca{}, which explains the \wsha{} correlation being weaker than that of the core-based \nsha{}.
In the case of HD\,20003 (middle panels), there is a strong anti-correlation between \sca{} and $S_{\text{H}\alpha\text{W}}$, which contributes to the mitigation of the strong positive correlation observed for \nsha{} in Fig. \ref{fig:example_ca_ha16_ha06} and results in a non-correlation for \wsha{}.
In the case of HD\,7199 (lower panels), the variation of flux with activity in the wings is higher (in negative values) than that of the core, as we saw in Fig. \ref{fig:ha_lines}, turning the positive correlation between \sca{} and \nsha{} observed in Fig. \ref{fig:example_ca_ha16_ha06} into a strong anti-correlation when the wings flux is added in \wsha{}.

Following these results, we can explain the correlation between \sha{} and \ha{} based on the balance between the flux in the core and that of the wings:
\begin{itemize}
	\item The flux in the \ha{} line core appears to be always (except for cases of low activity variability) positively correlated to that of the \cahk{} lines.
	\item The flux in the \ha{} line wings has a tendency to be negatively correlated to that of the \cahk{} lines., mainly in inactive and higher metallicity stars.
	\item When the flux in the \ha{} line wings is included in the derivation of the \sha{} index, there will be strong negative correlations with the \cahk{} lines for inactive and higher metallicity stars when the variation of flux with activity in the wings is stronger than that of the \ha{} core. For the more active stars (\rhk $> -4.8$ dex) the flux variation in the wings is irrelevant when compared to that of the core, producing a strong positive correlation.
\end{itemize}
Since the \cahk{} lines are known to be well correlated with the presence of plages/faculae this indicates that the core of the \ha{} line is following the same phenomena.
The existence of strong (negative) correlations between the flux in the \ha{} line wings and \cahk{} shows that the wings contain activity information.
\citet{meunier2009} discussed the contributions of plages and filaments to the flux in the \ha{} line, showing that the contribution of plages would produce a positive correlation between the flux in \ca{} and \ha{}, while filaments would contribute to a negative correlation, and thus, the combination of plages and filaments could result in non-correlations or anti-correlations depending on the plages and filaments filling factors, contrasts, and/or distributions \citep[see also][]{meunier2022}.
It seems that, by separating the \ha{} line core from the line wings, we could be observing those behaviours separately, meaning that the flux in the \ha{} wings could be following the presence of filaments in the stellar disk.
However, the identification of the structures responsible for the observed correlations is beyond the scope of the present work.
Further investigation could be carried out by obtaining high-resolution spectra of different resolved activity structures in the Sun to analyse the \ha{} profile and compare the flux in the \ha{} line, measured with different bandwidths with the presence of specific activity features.
This could be done e.g. with the future PoET Solar telescope (Santos et al. in prep.) to be installed at Paranal, Chile, and connected to the ESPRESSO spectrograph \citep{pepe2021}.

\section{Conclusions}\label{sec:conclusions}
In this work we analysed the effect of varying the \ha{} bandwidth on the correlation between the \sca{} and \sha{} activity indices using a sample of FGK stars observed with HARPS with cadence and long timespans enough to detect rotation and cycle induced variability.
We also compared the activity signals coming from long-term cycles and short-term rotation modulation using representative bandwidths: a narrrow (0.6 \AA{}), wide (1.6 \AA{}) bandpasses, and also a bandpass using just the line wings ($1.6 - 0.6$ \AA{}).

While studying the correlations between \sca{} and \sha{} for long- and short-timescales we found in general similar results. However the statistical significance is stronger for the long-timescales case, probably due to the increased number of data points used.

The effect of changing the \ha{} bandwidth for the correlation between \sca{} and \sha{} can be sumarised as follows:
\begin{itemize}
	\item Narrower bandwidths have a tendency for strong positive correlations with \sca{}.
	\item Wider bandwidths result in a great variety of correlations.
	\item There are no cases where the correlation is stronger when using a wider bandwidth than using a narrower.
	\item Narrow bandwidths can convert some strong negative correlations observed with wider bands to strong positive correlations.
	\item The bandwidth that maximises positive correlations with \sca{} is 0.6 \AA{}.
\end{itemize}

Previous works investigating the correlation between \sca{} and \sha{} for FGK dwarfs using wide bandwidths, generaly of 1.6 \AA{}, have found that the correlation between \sca{} and \sha{} depends on activity level and metallicity, where for low activity levels and higher metallicity there are non- and negative correlations and for higher activity levels the correlations are strong positive \citep{gomesdasilva2014, meunier2022}.
They also found that the correlation is not dependent on the effective temperature of stars.
In this work we arrived at the same results.
However, when analysing the correlations using a narrower bandpass of 0.6 \AA{} we found that most stars have strong positive correlations, independently of activity level and metallicity.
Furthermore, the cause for non correlations appears to be low activity variability.

We also analysed an index based on the flux of the \ha{} wings, and studied its correlation with \sca{}.
We found that the behaviour is similar to that of the \ha{} index with the wide 1.6 \AA{} bandpass, but without strong positive correlations and with more cases of strong negative correlations.

Regarding the time series, we observed that both long-term activity cycle variations and short-term rotational modulated variations are better followed by \sha{} if using the narrower 0.6 \AA{} bandwidth, mainly for inactive and K dwarfs.

In conclusion, the correlation between \sca{} and \wsha{} depends on the balance between the flux variation with activity between the line core and line wings of \ha{}.
While the \ha{} line core is positively correlated with \sca{}, the \ha{} wings show signs of following activity that is generally anti-correlated with the line core.
This kind of \ha{} profile behaviour was previously observed for two solar analogs by \citet{flores2016, flores2018} where they suggested that the \ha{} index should be constructed taking into account the profile variations instead of the integrated flux in the line.
Here, we suggest a simple alternative solution by integrating the flux in a narrower bandwidth covering only the line core, which is able to achieve similar results.

Having demonstrated that the widely used 1.6 \AA{} bandwidth on \ha{} degrades the quality of activity signals both at long and short timescales, we recommend the use of a narrower bandwidth, around 0.6 \AA{} or between $0.25 \geq \Delta \lambda \geq 0.75$ \AA{}, when calculating the \ha{} index to better identify both rotationally modulated and magnetic cycle activity signals and to decrease false positives when identifying activity with the \ha{} for radial-velocity exoplanet searches.

\begin{acknowledgements}
	This work was supported by FCT - Fundação para a Ciência e a Tecnologia through national funds and by FEDER through COMPETE2020 - Programa Operacional Competitividade e Internacionalização by these grants: UIDB/04434/2020 \& UIDP/04434/2020; PTDC/FIS-AST/32113/2017 \& POCI-01-0145-FEDER-032113; PTDC/FIS-AST/28953/2017 \& POCI-01-0145-FEDER-028953. 
\end{acknowledgements}

\bibliographystyle{aa} 
\bibliography{bibliography.bib}

\begin{appendix}

\FloatBarrier

\section{Plots of the $S_{\text{H}\alpha\text{W}}$ analysis}

\begin{figure}
	\resizebox{\hsize}{!}{\includegraphics{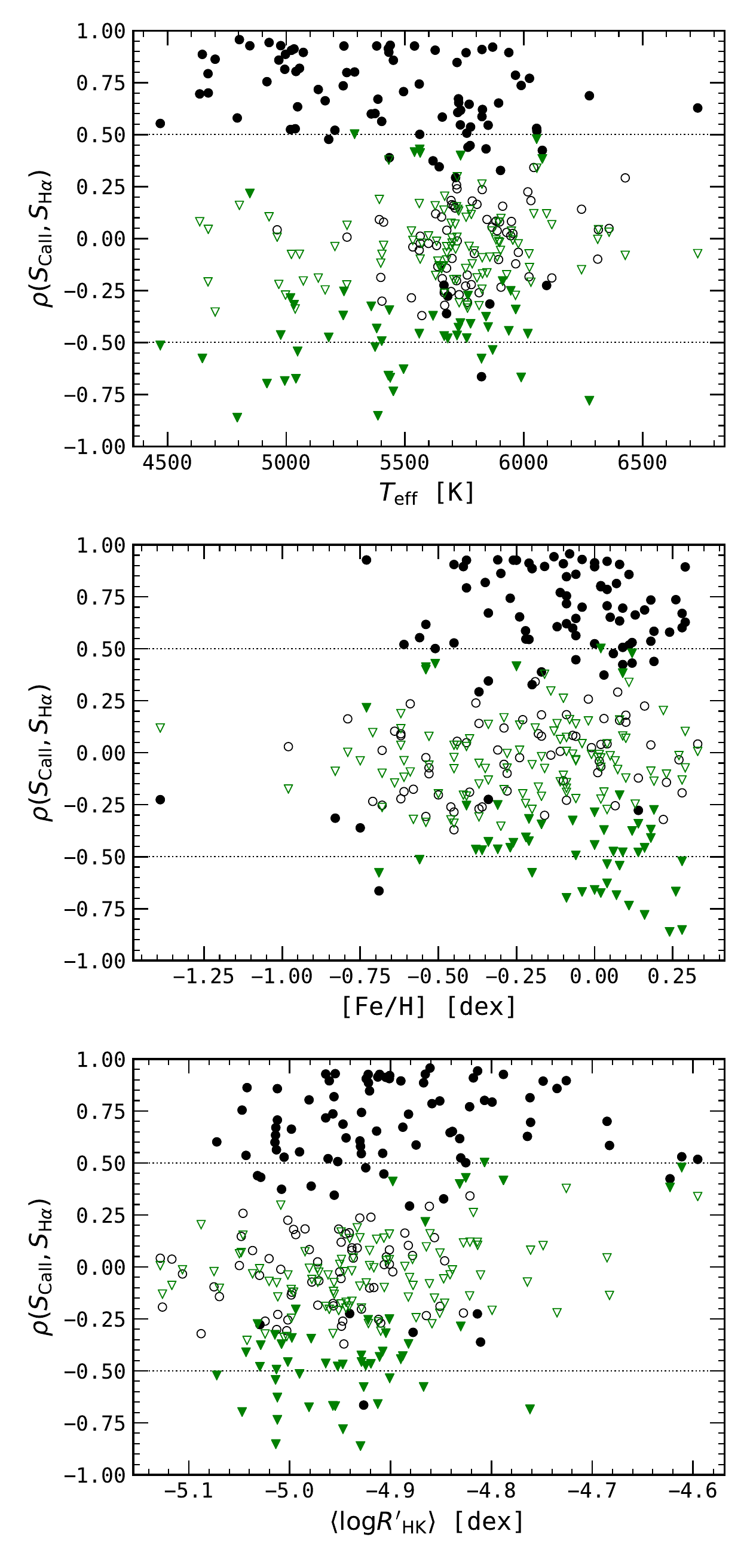}}

	\caption{Spearman correlation coefficient between \sca{} and \nsha{} (black circles) and \sca{} and $S_{\text{H}\alpha\text{W}}$ (green triangles) against effective temperature (top panel), metallicity (middle panel), and median \rhk{}.
	Horizontal dotted lines mark the thresholds for strong correlations at $\rho = 0.5$ and $-0.5$.
	Filled markers are significant correlations with $p$-value $\leq 10^{-3}$.
	}
	\label{fig:rho06_params_wings}
\end{figure}

\begin{figure*}
	\resizebox{\hsize}{!}{\includegraphics{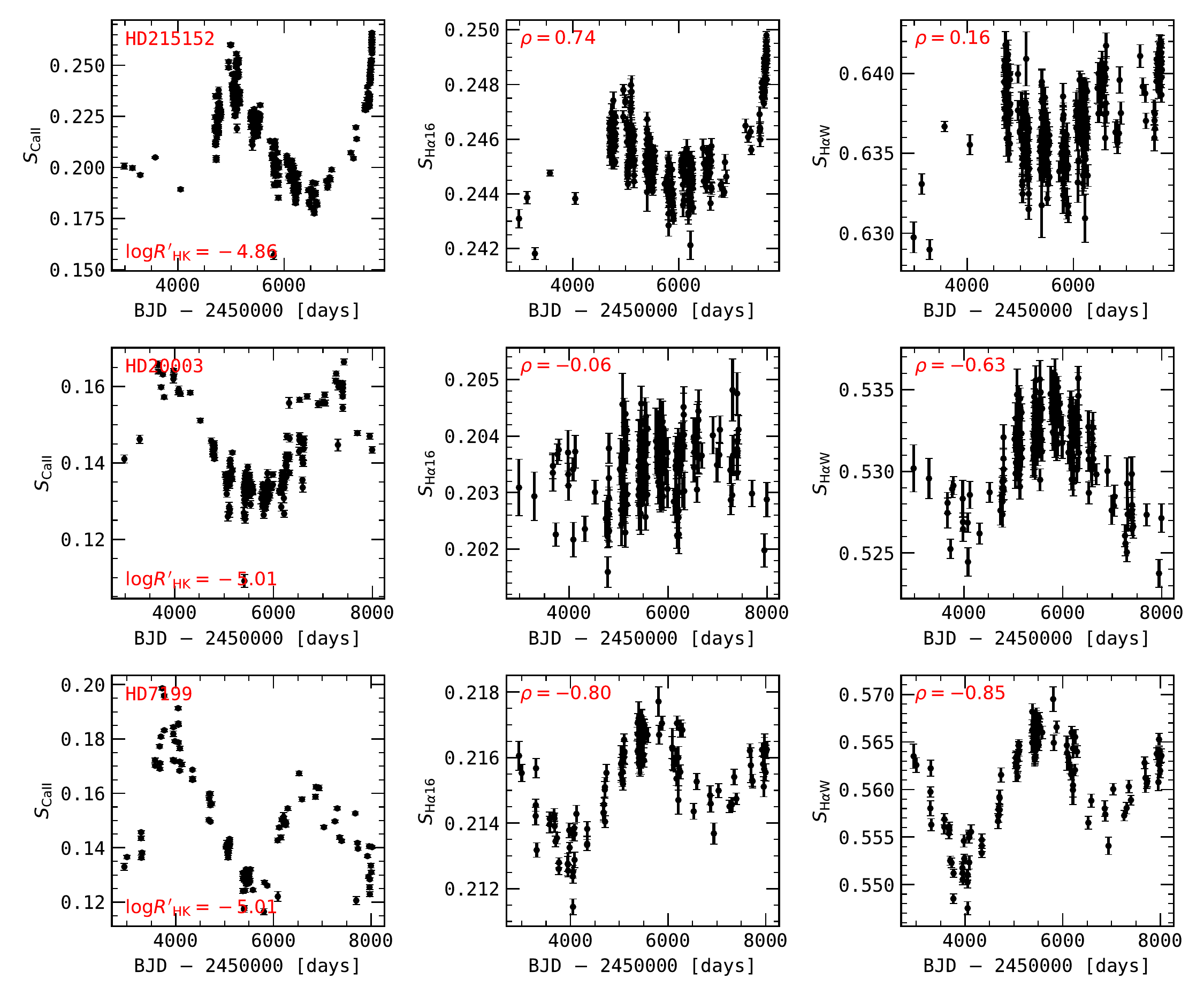}}

	\caption{\sca{}, \wsha{}, and $S_{\text{H}\alpha\text{W}}$ for HD\,215152 (top panels), HD\,20003 (middle panels), and HD\,7199 (lower panels).}

	\label{fig:example_ca_ha16_hawings}
\end{figure*}

\FloatBarrier

\section{\ca{} time series and selection of maxima and minima}\label{app:plots_ca_time_series}
Figure \ref{fig:ca_time_series} shows the \ca{} time series for the 103 stars used in the short-term analysis in \S \ref{sec:short_term}.
Red and blue points mark the subsets used to calculate correlations at the higher and lower levels of activity, respectively.

\begin{figure*}
    \setcounter{figure}{0}
	\resizebox{\hsize}{!}{\includegraphics{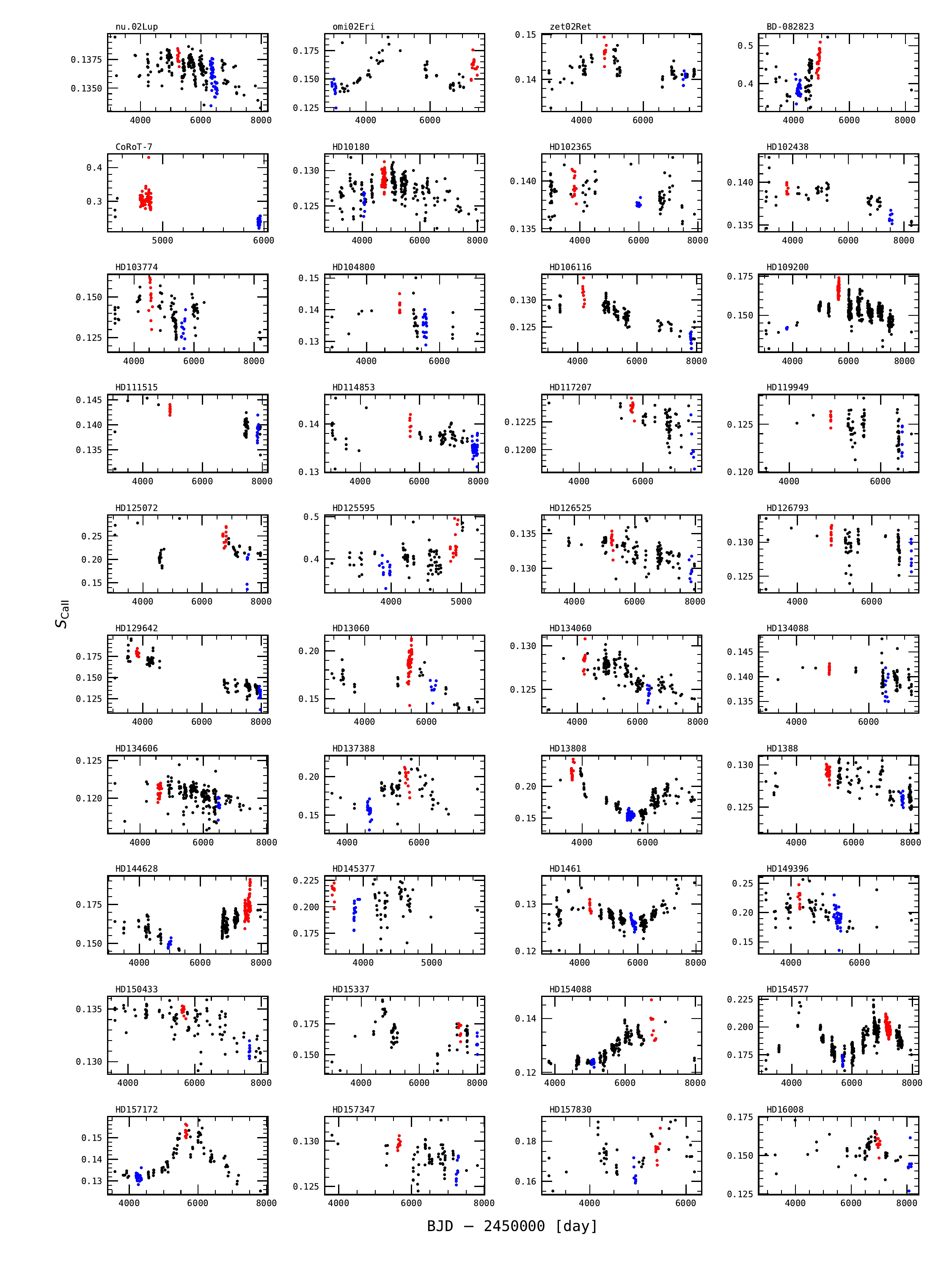}}

	\caption{
		\sca{} time series. Red points are the epoch at maximum, and blue points are the epoch at minimum.
	}
	\label{fig:ca_time_series}
\end{figure*}

\begin{figure*}
    \setcounter{figure}{0}
	\resizebox{\hsize}{!}{\includegraphics{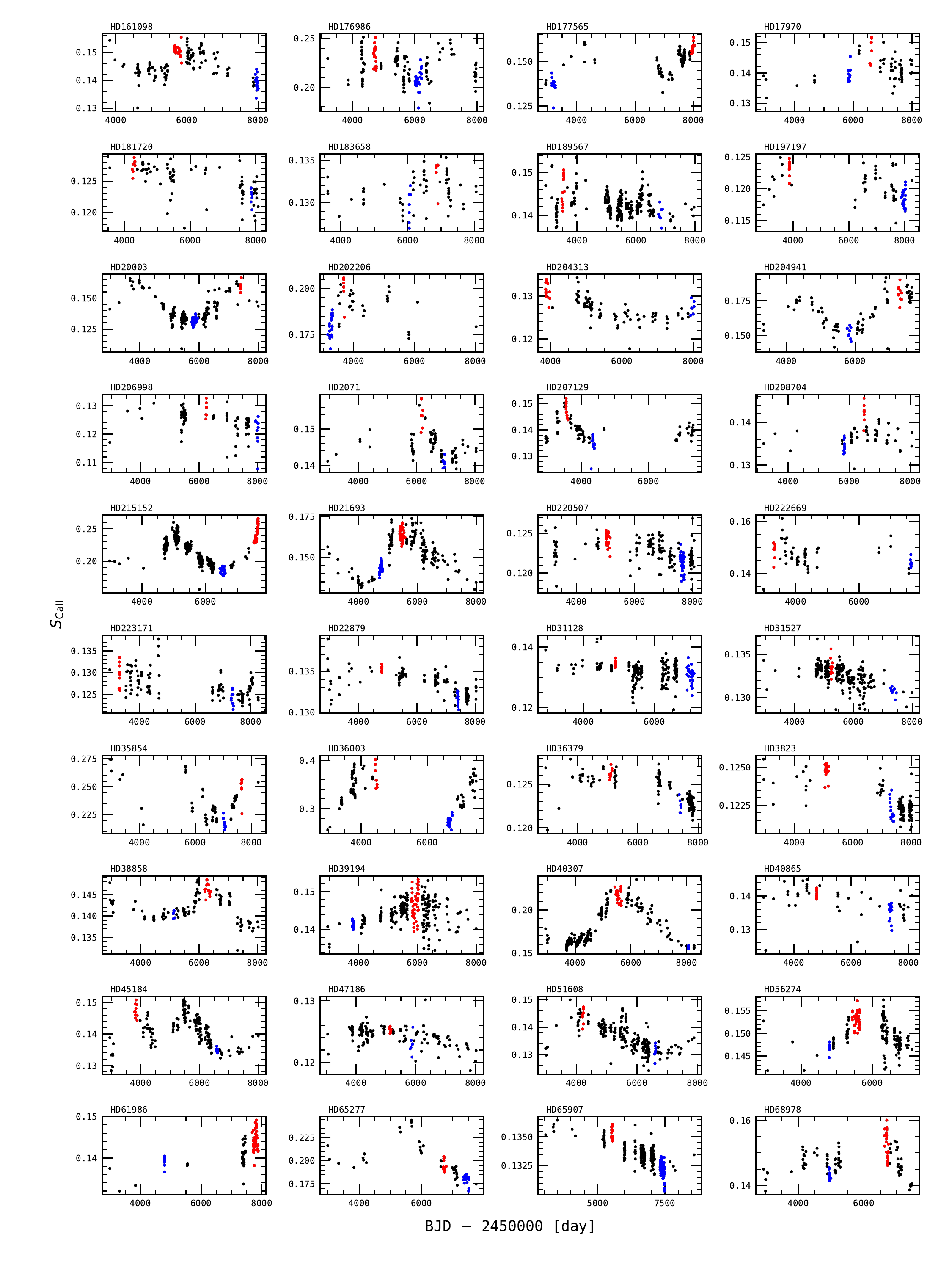}}

	\caption{
		continued.
	}
\end{figure*}

\begin{figure*}
    \setcounter{figure}{0}
	\resizebox{\hsize}{!}{\includegraphics{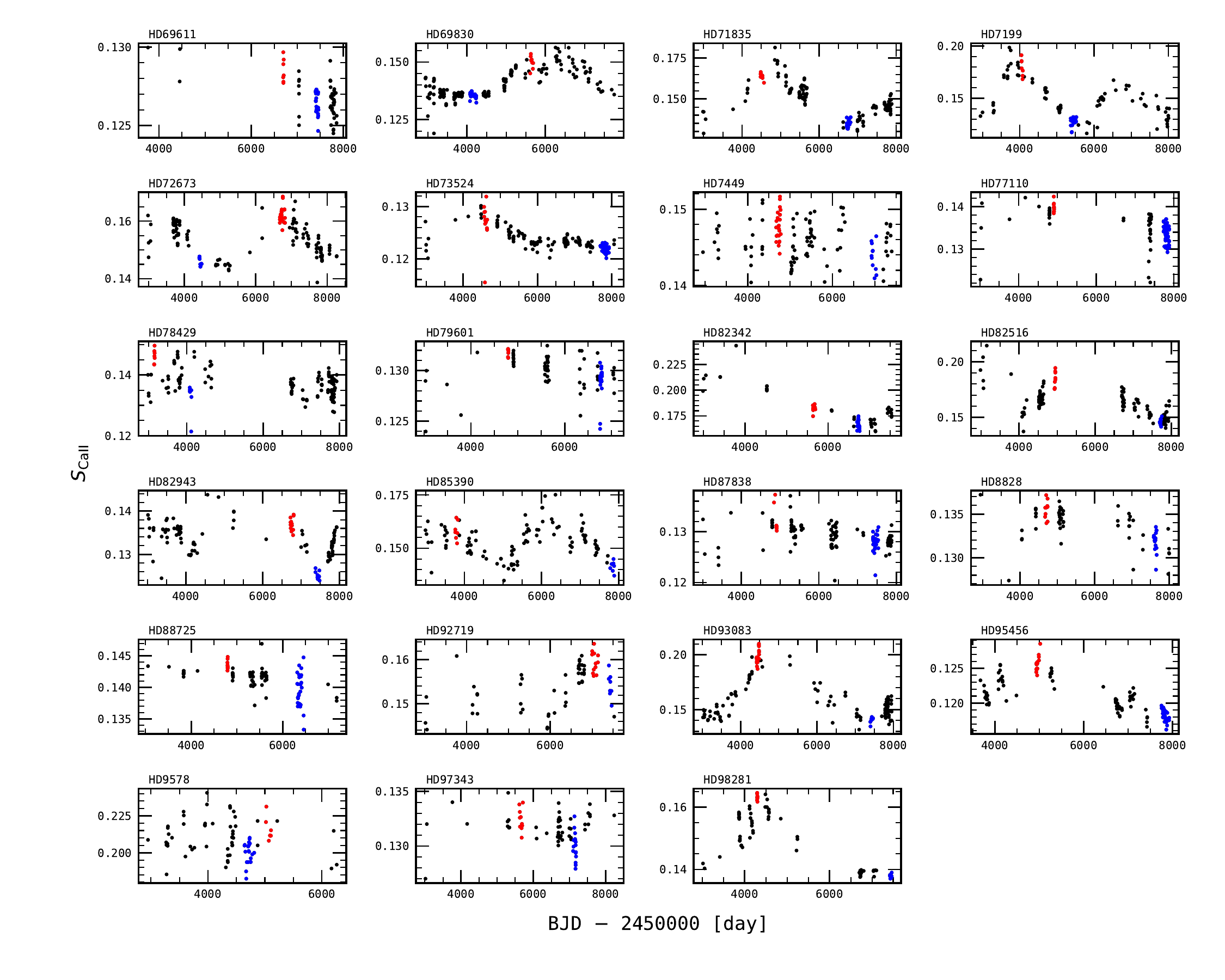}}

	\caption{
		continued.
	}
\end{figure*}

\FloatBarrier

\section{Tables of stellar parameters and correlations}

\longtab[1]{
\tiny
\begin{longtable}{lcccccc}

\caption{\label{tab:stellar_table}Stellar parameters for the sample.} \\

\hline\hline
\multicolumn{1}{l}{Star}&
\multicolumn{1}{c}{$N_\mathrm{nights}$}&
\multicolumn{1}{c}{$T_\mathrm{span}$}&
\multicolumn{1}{c}{Sp.Type}&
\multicolumn{1}{c}{$T_\mathrm{eff}$}&
\multicolumn{1}{c}{[Fe/H]}&
\multicolumn{1}{c}{$\langle \log R'_\mathrm{HK} \rangle$}
\\
& 
& 
[days] & 
& 
[K] & 
[dex] & 
[dex]
\\

\hline
\endfirsthead
\caption{continued.} \\
\hline\hline
\multicolumn{1}{l}{Star}&
\multicolumn{1}{c}{$N_\mathrm{nights}$}&
\multicolumn{1}{c}{$T_\mathrm{span}$}&
\multicolumn{1}{c}{Sp.Type}&
\multicolumn{1}{c}{$T_\mathrm{eff}$}&
\multicolumn{1}{c}{[Fe/H]}&
\multicolumn{1}{c}{$\langle \log R'_\mathrm{HK} \rangle$}
\\
& 
& 
[days] & 
& 
[K] & 
[dex] & 
[dex]
\\

\hline
\endhead
\hline
\endfoot
18\,Pup &70 &$2740$ &F5 &$6427$ $\pm$ $44$ &$0.07$ $\pm$ $0.03$ &$-4.862$ $\pm$ $0.004$ \\
nu.02\,Lup &246 &$4815$ &G5 &$5664$ $\pm$ $14$ &$-0.34$ $\pm$ $0.01$ &$-4.941$ $\pm$ $0.007$ \\
omi02\,Eri &96 &$4530$ &K2 &$5072$ $\pm$ $53$ &$-0.42$ $\pm$ $0.04$ &$-4.961$ $\pm$ $0.045$ \\
zet02\,Ret &99 &$4723$ &G2 &$5824$ $\pm$ $15$ &$-0.22$ $\pm$ $0.01$ &$-4.875$ $\pm$ $0.014$ \\
BD-08\,2823 &106 &$5209$ &K4 &$4672$ $\pm$ $104$ &$-0.04$ $\pm$ $0.06$ &$-4.685$ $\pm$ $0.039$ \\
CoRoT-7 &97 &$1437$ &K0 &$5288$ $\pm$ $27$ &$0.02$ $\pm$ $0.02$ &$-4.807$ $\pm$ $0.046$ \\
HD\,10180 &268 &$5048$ &G0 &$5911$ $\pm$ $19$ &$0.08$ $\pm$ $0.01$ &$-4.994$ $\pm$ $0.013$ \\
HD\,101930 &73 &$4528$ &K1 &$5164$ $\pm$ $61$ &$0.13$ $\pm$ $0.04$ &$-4.998$ $\pm$ $0.040$ \\
HD\,102117 &71 &$4509$ &G5 &$5657$ $\pm$ $24$ &$0.28$ $\pm$ $0.02$ &$-5.126$ $\pm$ $0.017$ \\
HD\,102365 &102 &$4934$ &G5 &$5629$ $\pm$ $29$ &$-0.29$ $\pm$ $0.02$ &$-4.949$ $\pm$ $0.009$ \\
HD\,102438 &66 &$5228$ &G7 &$5560$ $\pm$ $13$ &$-0.29$ $\pm$ $0.01$ &$-4.949$ $\pm$ $0.010$ \\
HD\,103774 &115 &$4850$ &F3 &$6732$ $\pm$ $56$ &$0.29$ $\pm$ $0.03$ &$-4.764$ $\pm$ $0.050$ \\
HD\,104067 &87 &$2271$ &K2 &$4969$ $\pm$ $72$ &$-0.06$ $\pm$ $0.05$ &$-4.735$ $\pm$ $0.025$ \\
HD\,104800 &62 &$3984$ &G4 &$5697$ $\pm$ $25$ &$-0.79$ $\pm$ $0.02$ &$-4.886$ $\pm$ $0.025$ \\
HD\,106116 &134 &$4888$ &G4 &$5680$ $\pm$ $15$ &$0.14$ $\pm$ $0.01$ &$-5.030$ $\pm$ $0.018$ \\
HD\,107094 &76 &$5155$ &G6 &$5562$ $\pm$ $17$ &$-0.51$ $\pm$ $0.01$ &$-4.825$ $\pm$ $0.028$ \\
HD\,109200 &458 &$5195$ &K2 &$5056$ $\pm$ $33$ &$-0.35$ $\pm$ $0.02$ &$-4.956$ $\pm$ $0.027$ \\
HD\,109271 &99 &$4807$ &G2 &$5783$ $\pm$ $18$ &$0.10$ $\pm$ $0.01$ &$-4.996$ $\pm$ $0.048$ \\
HD\,111515 &64 &$4876$ &G9 &$5398$ $\pm$ $18$ &$-0.61$ $\pm$ $0.01$ &$-4.957$ $\pm$ $0.014$ \\
HD\,111777 &107 &$4808$ &G5 &$5666$ $\pm$ $19$ &$-0.68$ $\pm$ $0.01$ &$-4.912$ $\pm$ $0.013$ \\
HD\,11397 &69 &$5091$ &G6 &$5564$ $\pm$ $26$ &$-0.54$ $\pm$ $0.02$ &$-4.898$ $\pm$ $0.021$ \\
HD\,114853 &127 &$4934$ &G3 &$5705$ $\pm$ $14$ &$-0.23$ $\pm$ $0.01$ &$-4.944$ $\pm$ $0.014$ \\
HD\,117207 &78 &$4565$ &G5 &$5667$ $\pm$ $21$ &$0.22$ $\pm$ $0.02$ &$-5.088$ $\pm$ $0.010$ \\
HD\,119173 &97 &$4400$ &G2 &$5779$ $\pm$ $44$ &$-0.62$ $\pm$ $0.03$ &$-4.828$ $\pm$ $0.018$ \\
HD\,119949 &82 &$3186$ &F6 &$6359$ $\pm$ $36$ &$-0.41$ $\pm$ $0.02$ &$-4.904$ $\pm$ $0.012$ \\
HD\,125072 &71 &$4919$ &K3 &$4794$ $\pm$ $102$ &$0.24$ $\pm$ $0.08$ &$-4.930$ $\pm$ $0.047$ \\
HD\,125595 &113 &$2075$ &K4 &$4636$ $\pm$ $84$ &$0.09$ $\pm$ $0.04$ &$-4.761$ $\pm$ $0.034$ \\
HD\,126525 &130 &$4814$ &G5 &$5638$ $\pm$ $13$ &$-0.10$ $\pm$ $0.01$ &$-4.999$ $\pm$ $0.012$ \\
HD\,126793 &85 &$3913$ &G0 &$5904$ $\pm$ $33$ &$-0.71$ $\pm$ $0.02$ &$-4.865$ $\pm$ $0.013$ \\
HD\,129642 &129 &$4914$ &K3 &$4919$ $\pm$ $65$ &$-0.09$ $\pm$ $0.03$ &$-5.047$ $\pm$ $0.068$ \\
HD\,13060 &90 &$4702$ &K0 &$5255$ $\pm$ $45$ &$0.02$ $\pm$ $0.03$ &$-4.851$ $\pm$ $0.065$ \\
HD\,133633 &68 &$3076$ &G6 &$5571$ $\pm$ $19$ &$-0.45$ $\pm$ $0.01$ &$-4.946$ $\pm$ $0.023$ \\
HD\,134060 &159 &$4822$ &G0 &$5966$ $\pm$ $14$ &$0.14$ $\pm$ $0.01$ &$-4.998$ $\pm$ $0.012$ \\
HD\,134088 &86 &$4033$ &G4 &$5675$ $\pm$ $22$ &$-0.75$ $\pm$ $0.02$ &$-4.811$ $\pm$ $0.013$ \\
HD\,134606 &217 &$4593$ &G5 &$5633$ $\pm$ $28$ &$0.27$ $\pm$ $0.02$ &$-5.106$ $\pm$ $0.010$ \\
HD\,136713 &70 &$4152$ &K2 &$4994$ $\pm$ $74$ &$0.07$ $\pm$ $0.05$ &$-4.762$ $\pm$ $0.036$ \\
HD\,137388 &92 &$4037$ &K0 &$5240$ $\pm$ $53$ &$0.18$ $\pm$ $0.03$ &$-4.882$ $\pm$ $0.054$ \\
HD\,13808 &228 &$4438$ &K2 &$5033$ $\pm$ $38$ &$-0.21$ $\pm$ $0.02$ &$-4.905$ $\pm$ $0.075$ \\
HD\,1388 &147 &$5082$ &G0 &$5954$ $\pm$ $10$ &$-0.01$ $\pm$ $0.01$ &$-4.972$ $\pm$ $0.012$ \\
HD\,144628 &265 &$4767$ &K2 &$5022$ $\pm$ $26$ &$-0.45$ $\pm$ $0.02$ &$-4.924$ $\pm$ $0.027$ \\
HD\,145377 &75 &$2122$ &F9 &$6054$ $\pm$ $16$ &$0.12$ $\pm$ $0.01$ &$-4.611$ $\pm$ $0.035$ \\
HD\,1461 &255 &$5008$ &G2 &$5765$ $\pm$ $18$ &$0.19$ $\pm$ $0.01$ &$-5.032$ $\pm$ $0.013$ \\
HD\,148211 &70 &$4335$ &G0 &$5948$ $\pm$ $22$ &$-0.62$ $\pm$ $0.02$ &$-4.901$ $\pm$ $0.014$ \\
HD\,149396 &104 &$4249$ &G5 &$5657$ $\pm$ $23$ &$0.19$ $\pm$ $0.02$ &$-4.683$ $\pm$ $0.051$ \\
HD\,150433 &112 &$4366$ &G5 &$5665$ $\pm$ $12$ &$-0.36$ $\pm$ $0.01$ &$-4.947$ $\pm$ $0.011$ \\
HD\,15337 &83 &$5014$ &K1 &$5179$ $\pm$ $44$ &$0.06$ $\pm$ $0.03$ &$-4.925$ $\pm$ $0.038$ \\
HD\,154088 &186 &$4143$ &G9 &$5374$ $\pm$ $43$ &$0.28$ $\pm$ $0.03$ &$-5.072$ $\pm$ $0.025$ \\
HD\,154577 &443 &$4826$ &K3 &$4847$ $\pm$ $35$ &$-0.73$ $\pm$ $0.01$ &$-4.866$ $\pm$ $0.030$ \\
HD\,157172 &123 &$4241$ &G8 &$5451$ $\pm$ $27$ &$0.11$ $\pm$ $0.02$ &$-5.012$ $\pm$ $0.042$ \\
HD\,157347 &85 &$3998$ &G4 &$5676$ $\pm$ $16$ &$0.02$ $\pm$ $0.01$ &$-5.020$ $\pm$ $0.011$ \\
HD\,157830 &63 &$3024$ &G7 &$5540$ $\pm$ $16$ &$-0.25$ $\pm$ $0.01$ &$-4.788$ $\pm$ $0.033$ \\
HD\,16008 &76 &$5215$ &G2 &$5770$ $\pm$ $14$ &$-0.06$ $\pm$ $0.01$ &$-4.841$ $\pm$ $0.034$ \\
HD\,161098 &149 &$4172$ &G7 &$5560$ $\pm$ $15$ &$-0.27$ $\pm$ $0.01$ &$-4.929$ $\pm$ $0.026$ \\
HD\,172568 &69 &$3795$ &G3 &$5728$ $\pm$ $22$ &$-0.37$ $\pm$ $0.02$ &$-4.910$ $\pm$ $0.020$ \\
HD\,175607 &111 &$3912$ &G9 &$5392$ $\pm$ $17$ &$-0.62$ $\pm$ $0.01$ &$-4.933$ $\pm$ $0.020$ \\
HD\,176986 &161 &$4762$ &K2 &$5018$ $\pm$ $59$ &$0.00$ $\pm$ $0.03$ &$-4.830$ $\pm$ $0.028$ \\
HD\,177565 &145 &$5082$ &G5 &$5627$ $\pm$ $19$ &$0.08$ $\pm$ $0.01$ &$-4.901$ $\pm$ $0.036$ \\
HD\,17970 &74 &$5063$ &K2 &$5038$ $\pm$ $31$ &$-0.45$ $\pm$ $0.02$ &$-5.006$ $\pm$ $0.019$ \\
HD\,181433 &184 &$5029$ &K2 &$4962$ $\pm$ $134$ &$0.33$ $\pm$ $0.13$ &$-5.128$ $\pm$ $0.016$ \\
HD\,181720 &89 &$4519$ &G2 &$5792$ $\pm$ $17$ &$-0.53$ $\pm$ $0.01$ &$-4.978$ $\pm$ $0.019$ \\
HD\,183658 &72 &$4446$ &G2 &$5803$ $\pm$ $17$ &$0.03$ $\pm$ $0.01$ &$-4.941$ $\pm$ $0.014$ \\
HD\,189567 &256 &$5031$ &G3 &$5726$ $\pm$ $15$ &$-0.24$ $\pm$ $0.01$ &$-4.914$ $\pm$ $0.016$ \\
HD\,19467 &71 &$5110$ &G3 &$5720$ $\pm$ $10$ &$-0.14$ $\pm$ $0.01$ &$-5.009$ $\pm$ $0.009$ \\
HD\,197027 &66 &$3077$ &G4 &$5694$ $\pm$ $28$ &$0.07$ $\pm$ $0.02$ &$-5.002$ $\pm$ $0.020$ \\
HD\,197197 &78 &$5313$ &G2 &$5812$ $\pm$ $16$ &$-0.46$ $\pm$ $0.01$ &$-5.024$ $\pm$ $0.020$ \\
HD\,199289 &64 &$4012$ &G0 &$5928$ $\pm$ $37$ &$-0.98$ $\pm$ $0.03$ &$-4.846$ $\pm$ $0.013$ \\
HD\,199847 &81 &$5120$ &G2 &$5763$ $\pm$ $20$ &$-0.54$ $\pm$ $0.02$ &$-5.003$ $\pm$ $0.025$ \\
HD\,20003 &193 &$5010$ &G8 &$5494$ $\pm$ $27$ &$0.04$ $\pm$ $0.02$ &$-5.012$ $\pm$ $0.057$ \\
HD\,202206 &62 &$4872$ &G2 &$5757$ $\pm$ $25$ &$0.29$ $\pm$ $0.02$ &$-4.749$ $\pm$ $0.035$ \\
HD\,20407 &72 &$5130$ &G1 &$5866$ $\pm$ $14$ &$-0.44$ $\pm$ $0.01$ &$-4.878$ $\pm$ $0.009$ \\
HD\,204313 &106 &$4158$ &G2 &$5776$ $\pm$ $22$ &$0.18$ $\pm$ $0.02$ &$-5.043$ $\pm$ $0.020$ \\
HD\,204941 &96 &$4312$ &K2 &$4997$ $\pm$ $36$ &$-0.20$ $\pm$ $0.02$ &$-4.922$ $\pm$ $0.041$ \\
HD\,206998 &93 &$5076$ &G2 &$5822$ $\pm$ $26$ &$-0.69$ $\pm$ $0.02$ &$-4.927$ $\pm$ $0.025$ \\
HD\,2071 &83 &$5104$ &G3 &$5719$ $\pm$ $14$ &$-0.09$ $\pm$ $0.01$ &$-4.921$ $\pm$ $0.021$ \\
HD\,207129 &107 &$4418$ &G0 &$5937$ $\pm$ $13$ &$0.00$ $\pm$ $0.01$ &$-4.890$ $\pm$ $0.027$ \\
HD\,20781 &224 &$5014$ &K0 &$5256$ $\pm$ $29$ &$-0.11$ $\pm$ $0.02$ &$-5.050$ $\pm$ $0.013$ \\
HD\,20782 &84 &$5109$ &G2 &$5774$ $\pm$ $14$ &$-0.06$ $\pm$ $0.01$ &$-4.907$ $\pm$ $0.015$ \\
HD\,207869 &80 &$5034$ &G7 &$5527$ $\pm$ $21$ &$-0.45$ $\pm$ $0.02$ &$-4.949$ $\pm$ $0.027$ \\
HD\,208704 &69 &$4846$ &G1 &$5826$ $\pm$ $11$ &$-0.09$ $\pm$ $0.01$ &$-4.944$ $\pm$ $0.017$ \\
HD\,210918 &136 &$5108$ &G2 &$5755$ $\pm$ $12$ &$-0.09$ $\pm$ $0.01$ &$-5.012$ $\pm$ $0.009$ \\
HD\,21132 &92 &$4245$ &F7 &$6243$ $\pm$ $34$ &$-0.37$ $\pm$ $0.02$ &$-4.856$ $\pm$ $0.035$ \\
HD\,211415 &65 &$4433$ &G1 &$5850$ $\pm$ $14$ &$-0.21$ $\pm$ $0.01$ &$-4.929$ $\pm$ $0.016$ \\
HD\,21209 &64 &$4720$ &K4 &$4671$ $\pm$ $65$ &$-0.41$ $\pm$ $0.04$ &$-4.799$ $\pm$ $0.025$ \\
HD\,215152 &323 &$4671$ &K3 &$4803$ $\pm$ $52$ &$-0.08$ $\pm$ $0.02$ &$-4.861$ $\pm$ $0.047$ \\
HD\,21693 &218 &$5060$ &G8 &$5430$ $\pm$ $26$ &$0.00$ $\pm$ $0.02$ &$-4.913$ $\pm$ $0.050$ \\
HD\,219828 &91 &$3751$ &G1 &$5888$ $\pm$ $14$ &$0.18$ $\pm$ $0.01$ &$-5.117$ $\pm$ $0.018$ \\
HD\,220507 &199 &$5109$ &G4 &$5698$ $\pm$ $17$ &$0.01$ $\pm$ $0.01$ &$-5.075$ $\pm$ $0.012$ \\
HD\,222669 &68 &$4677$ &G1 &$5894$ $\pm$ $17$ &$0.05$ $\pm$ $0.01$ &$-4.839$ $\pm$ $0.021$ \\
HD\,223171 &134 &$5327$ &G1 &$5841$ $\pm$ $18$ &$0.12$ $\pm$ $0.01$ &$-5.029$ $\pm$ $0.021$ \\
HD\,224817 &106 &$4266$ &G1 &$5894$ $\pm$ $22$ &$-0.53$ $\pm$ $0.02$ &$-4.921$ $\pm$ $0.014$ \\
HD\,22879 &146 &$5087$ &G1 &$5857$ $\pm$ $27$ &$-0.83$ $\pm$ $0.02$ &$-4.878$ $\pm$ $0.010$ \\
HD\,31103 &72 &$4770$ &F9 &$6078$ $\pm$ $16$ &$0.09$ $\pm$ $0.01$ &$-4.623$ $\pm$ $0.039$ \\
HD\,31128 &210 &$4170$ &F8 &$6096$ $\pm$ $67$ &$-1.39$ $\pm$ $0.04$ &$-4.814$ $\pm$ $0.016$ \\
HD\,31527 &248 &$5053$ &G1 &$5898$ $\pm$ $13$ &$-0.17$ $\pm$ $0.01$ &$-4.938$ $\pm$ $0.008$ \\
HD\,31822 &82 &$4523$ &F9 &$6042$ $\pm$ $16$ &$-0.19$ $\pm$ $0.01$ &$-4.821$ $\pm$ $0.010$ \\
HD\,32564 &189 &$2301$ &G7 &$5533$ $\pm$ $29$ &$0.01$ $\pm$ $0.02$ &$-5.030$ $\pm$ $0.015$ \\
HD\,35854 &77 &$5298$ &K3 &$4928$ $\pm$ $56$ &$-0.13$ $\pm$ $0.03$ &$-4.814$ $\pm$ $0.034$ \\
HD\,36003 &111 &$4489$ &K4 &$4647$ $\pm$ $88$ &$-0.20$ $\pm$ $0.06$ &$-4.867$ $\pm$ $0.050$ \\
HD\,36379 &147 &$4926$ &F9 &$6030$ $\pm$ $14$ &$-0.17$ $\pm$ $0.01$ &$-4.951$ $\pm$ $0.014$ \\
HD\,3823 &137 &$5081$ &F9 &$6022$ $\pm$ $14$ &$-0.28$ $\pm$ $0.01$ &$-4.972$ $\pm$ $0.009$ \\
HD\,38858 &119 &$5076$ &G3 &$5733$ $\pm$ $12$ &$-0.22$ $\pm$ $0.01$ &$-4.908$ $\pm$ $0.018$ \\
HD\,39194 &271 &$5050$ &K1 &$5205$ $\pm$ $23$ &$-0.61$ $\pm$ $0.02$ &$-4.962$ $\pm$ $0.017$ \\
HD\,40307 &253 &$5324$ &K2 &$4977$ $\pm$ $59$ &$-0.31$ $\pm$ $0.03$ &$-4.964$ $\pm$ $0.066$ \\
HD\,40865 &73 &$5174$ &G3 &$5719$ $\pm$ $16$ &$-0.38$ $\pm$ $0.01$ &$-4.919$ $\pm$ $0.021$ \\
HD\,41248 &163 &$5175$ &G3 &$5713$ $\pm$ $21$ &$-0.37$ $\pm$ $0.01$ &$-4.881$ $\pm$ $0.021$ \\
HD\,4308 &145 &$5081$ &G5 &$5644$ $\pm$ $16$ &$-0.34$ $\pm$ $0.01$ &$-4.956$ $\pm$ $0.013$ \\
HD\,45184 &189 &$5061$ &G1 &$5869$ $\pm$ $14$ &$0.04$ $\pm$ $0.01$ &$-4.901$ $\pm$ $0.030$ \\
HD\,45289 &62 &$4869$ &G3 &$5717$ $\pm$ $18$ &$-0.02$ $\pm$ $0.01$ &$-5.046$ $\pm$ $0.008$ \\
HD\,45364 &112 &$4866$ &G8 &$5434$ $\pm$ $20$ &$-0.17$ $\pm$ $0.01$ &$-4.979$ $\pm$ $0.022$ \\
HD\,47186 &145 &$4966$ &G4 &$5675$ $\pm$ $21$ &$0.23$ $\pm$ $0.02$ &$-5.070$ $\pm$ $0.010$ \\
HD\,51608 &217 &$4899$ &G9 &$5358$ $\pm$ $22$ &$-0.07$ $\pm$ $0.01$ &$-5.015$ $\pm$ $0.028$ \\
HD\,5388 &75 &$5067$ &F6 &$6311$ $\pm$ $33$ &$-0.28$ $\pm$ $0.02$ &$-4.917$ $\pm$ $0.039$ \\
HD\,56274 &171 &$4171$ &G3 &$5734$ $\pm$ $22$ &$-0.54$ $\pm$ $0.02$ &$-4.831$ $\pm$ $0.017$ \\
HD\,564 &106 &$5172$ &G0 &$5902$ $\pm$ $14$ &$-0.20$ $\pm$ $0.01$ &$-4.847$ $\pm$ $0.052$ \\
HD\,59468 &163 &$5301$ &G6 &$5618$ $\pm$ $20$ &$0.03$ $\pm$ $0.01$ &$-5.008$ $\pm$ $0.011$ \\
HD\,59711 &68 &$5144$ &G3 &$5722$ $\pm$ $13$ &$-0.12$ $\pm$ $0.01$ &$-4.930$ $\pm$ $0.014$ \\
HD\,61986 &95 &$4888$ &G3 &$5725$ $\pm$ $20$ &$-0.34$ $\pm$ $0.02$ &$-4.888$ $\pm$ $0.018$ \\
HD\,63765 &68 &$5241$ &G8 &$5432$ $\pm$ $19$ &$-0.16$ $\pm$ $0.01$ &$-4.726$ $\pm$ $0.046$ \\
HD\,65277 &69 &$4767$ &K4 &$4701$ $\pm$ $57$ &$-0.30$ $\pm$ $0.04$ &$-5.042$ $\pm$ $0.041$ \\
HD\,65907 &301 &$5531$ &G0 &$5945$ $\pm$ $16$ &$-0.31$ $\pm$ $0.01$ &$-4.901$ $\pm$ $0.008$ \\
HD\,68978 &112 &$4525$ &G0 &$5965$ $\pm$ $22$ &$0.04$ $\pm$ $0.02$ &$-4.859$ $\pm$ $0.025$ \\
HD\,69611 &73 &$4094$ &G2 &$5762$ $\pm$ $25$ &$-0.58$ $\pm$ $0.02$ &$-4.957$ $\pm$ $0.011$ \\
HD\,69830 &264 &$4817$ &G9 &$5402$ $\pm$ $28$ &$-0.06$ $\pm$ $0.02$ &$-5.013$ $\pm$ $0.029$ \\
HD\,71334 &61 &$4822$ &G4 &$5694$ $\pm$ $13$ &$-0.09$ $\pm$ $0.01$ &$-4.985$ $\pm$ $0.010$ \\
HD\,71835 &150 &$4899$ &G8 &$5438$ $\pm$ $22$ &$-0.04$ $\pm$ $0.02$ &$-4.955$ $\pm$ $0.053$ \\
HD\,7199 &125 &$5075$ &G9 &$5386$ $\pm$ $45$ &$0.28$ $\pm$ $0.03$ &$-5.014$ $\pm$ $0.088$ \\
HD\,72673 &158 &$5281$ &K0 &$5243$ $\pm$ $22$ &$-0.41$ $\pm$ $0.01$ &$-4.922$ $\pm$ $0.027$ \\
HD\,73524 &156 &$5082$ &F9 &$6017$ $\pm$ $13$ &$0.16$ $\pm$ $0.01$ &$-5.002$ $\pm$ $0.019$ \\
HD\,7449 &116 &$4452$ &F9 &$6024$ $\pm$ $13$ &$-0.11$ $\pm$ $0.01$ &$-4.822$ $\pm$ $0.015$ \\
HD\,77110 &101 &$4864$ &G3 &$5717$ $\pm$ $20$ &$-0.50$ $\pm$ $0.02$ &$-4.929$ $\pm$ $0.021$ \\
HD\,78429 &136 &$4940$ &G2 &$5760$ $\pm$ $19$ &$0.09$ $\pm$ $0.01$ &$-4.952$ $\pm$ $0.029$ \\
HD\,79601 &85 &$4023$ &G2 &$5825$ $\pm$ $25$ &$-0.59$ $\pm$ $0.02$ &$-4.931$ $\pm$ $0.010$ \\
HD\,82342 &65 &$4549$ &K5 &$4470$ $\pm$ $21$ &$-0.56$ $\pm$ $0.10$ &$-4.990$ $\pm$ $0.040$ \\
HD\,82516 &134 &$4959$ &K2 &$5041$ $\pm$ $57$ &$0.02$ $\pm$ $0.03$ &$-4.981$ $\pm$ $0.043$ \\
HD\,82943 &120 &$4910$ &F9 &$5989$ $\pm$ $20$ &$0.26$ $\pm$ $0.01$ &$-4.957$ $\pm$ $0.027$ \\
HD\,85390 &113 &$4893$ &K1 &$5135$ $\pm$ $45$ &$-0.09$ $\pm$ $0.02$ &$-4.964$ $\pm$ $0.029$ \\
HD\,87838 &152 &$4872$ &F8 &$6118$ $\pm$ $33$ &$-0.40$ $\pm$ $0.02$ &$-4.851$ $\pm$ $0.015$ \\
HD\,8828 &79 &$5058$ &G9 &$5403$ $\pm$ $25$ &$-0.16$ $\pm$ $0.02$ &$-5.013$ $\pm$ $0.012$ \\
HD\,88725 &94 &$4130$ &G5 &$5654$ $\pm$ $17$ &$-0.64$ $\pm$ $0.01$ &$-4.882$ $\pm$ $0.013$ \\
HD\,89839 &79 &$5147$ &F6 &$6314$ $\pm$ $24$ &$0.04$ $\pm$ $0.02$ &$-4.937$ $\pm$ $0.049$ \\
HD\,90156 &127 &$4925$ &G6 &$5599$ $\pm$ $12$ &$-0.24$ $\pm$ $0.01$ &$-4.950$ $\pm$ $0.006$ \\
HD\,92719 &67 &$4509$ &G2 &$5824$ $\pm$ $16$ &$-0.10$ $\pm$ $0.01$ &$-4.818$ $\pm$ $0.026$ \\
HD\,93083 &134 &$4937$ &K2 &$5048$ $\pm$ $66$ &$0.08$ $\pm$ $0.04$ &$-5.014$ $\pm$ $0.066$ \\
HD\,93385 &235 &$4882$ &G0 &$5977$ $\pm$ $18$ &$0.02$ $\pm$ $0.01$ &$-4.971$ $\pm$ $0.008$ \\
HD\,95456 &124 &$4250$ &F7 &$6276$ $\pm$ $22$ &$0.16$ $\pm$ $0.02$ &$-4.947$ $\pm$ $0.022$ \\
HD\,9578 &77 &$3314$ &F9 &$6055$ $\pm$ $14$ &$0.11$ $\pm$ $0.01$ &$-4.595$ $\pm$ $0.037$ \\
HD\,96423 &72 &$4756$ &G3 &$5711$ $\pm$ $18$ &$0.10$ $\pm$ $0.01$ &$-5.048$ $\pm$ $0.008$ \\
HD\,967 &73 &$5081$ &G6 &$5564$ $\pm$ $16$ &$-0.68$ $\pm$ $0.01$ &$-4.906$ $\pm$ $0.015$ \\
HD\,96700 &231 &$4926$ &G1 &$5845$ $\pm$ $13$ &$-0.18$ $\pm$ $0.01$ &$-4.938$ $\pm$ $0.008$ \\
HD\,97037 &63 &$4791$ &G1 &$5883$ $\pm$ $14$ &$-0.07$ $\pm$ $0.01$ &$-4.981$ $\pm$ $0.009$ \\
HD\,97343 &73 &$5225$ &G9 &$5410$ $\pm$ $20$ &$-0.06$ $\pm$ $0.01$ &$-5.037$ $\pm$ $0.008$ \\
\end{longtable}
\tablefoot{
$N_\text{nights}$ is the number of binned nights of observation, $T_\text{span}$ the number of days between first and last observation, Sp.Type is the spectral type, $T_\text{eff}$ the effective temperature, [Fe/H] the metallicity, $\langle \log R'_\text{HK} \rangle$ the median activity levels where the errors are standard deviations of the time series.
}
}

\longtab[2]{
\tiny
\begin{longtable}{lccccccc}

\caption{\label{tab:corr_table_long}Spearman correlation coefficients for the long-term dataset.} \\

\hline\hline
\multicolumn{1}{l}{Star}&
\multicolumn{1}{c}{$\rho^{06}$}&
\multicolumn{1}{c}{$p$-value$^{06}$}&
\multicolumn{1}{c}{$\rho^{16}$}&
\multicolumn{1}{c}{$p$-value$^{16}$}&
\multicolumn{1}{c}{$\rho^{W}$}&
\multicolumn{1}{c}{$p$-value$^{W}$}
\\

\hline
\endfirsthead
\caption{continued.} \\
\hline\hline
\multicolumn{1}{l}{Star}&
\multicolumn{1}{c}{$\rho^{06}$}&
\multicolumn{1}{c}{$p$-value$^{06}$}&
\multicolumn{1}{c}{$\rho^{16}$}&
\multicolumn{1}{c}{$p$-value$^{16}$}&
\multicolumn{1}{c}{$\rho^{W}$}&
\multicolumn{1}{c}{$p$-value$^{W}$}
\\

\hline
\endhead
\hline
\endfoot
18\,Pup &$0.29$ &$0.007$ &$0.11$ &$0.178$ &$-0.08$ &$0.248$ \\
nu.02\,Lup &$-0.22$ &$0.000$ &$0.09$ &$0.086$ &$0.14$ &$0.016$ \\
omi02\,Eri &$0.90$ &$0.000$ &$0.55$ &$0.000$ &$-0.20$ &$0.024$ \\
zet02\,Ret &$0.59$ &$0.000$ &$0.12$ &$0.125$ &$-0.24$ &$0.008$ \\
BD-08\,2823 &$0.70$ &$0.000$ &$0.43$ &$0.000$ &$0.04$ &$0.328$ \\
CoRoT-7 &$0.80$ &$0.000$ &$0.73$ &$0.000$ &$0.50$ &$0.000$ \\
HD\,10180 &$0.16$ &$0.005$ &$-0.14$ &$0.009$ &$-0.20$ &$0.000$ \\
HD\,101930 &$0.66$ &$0.000$ &$0.01$ &$0.465$ &$-0.25$ &$0.018$ \\
HD\,102117 &$-0.19$ &$0.052$ &$-0.23$ &$0.027$ &$-0.13$ &$0.140$ \\
HD\,102365 &$0.12$ &$0.118$ &$-0.13$ &$0.089$ &$-0.18$ &$0.036$ \\
HD\,102438 &$-0.06$ &$0.326$ &$0.17$ &$0.086$ &$0.17$ &$0.087$ \\
HD\,103774 &$0.63$ &$0.000$ &$0.52$ &$0.000$ &$-0.07$ &$0.222$ \\
HD\,104067 &$0.86$ &$0.000$ &$0.26$ &$0.008$ &$-0.22$ &$0.020$ \\
HD\,104800 &$0.16$ &$0.103$ &$0.05$ &$0.354$ &$0.00$ &$0.493$ \\
HD\,106116 &$-0.28$ &$0.001$ &$-0.50$ &$0.000$ &$-0.48$ &$0.000$ \\
HD\,107094 &$0.50$ &$0.000$ &$0.58$ &$0.000$ &$0.43$ &$0.000$ \\
HD\,109200 &$0.82$ &$0.000$ &$0.38$ &$0.000$ &$-0.08$ &$0.053$ \\
HD\,109271 &$0.18$ &$0.036$ &$-0.05$ &$0.321$ &$-0.12$ &$0.113$ \\
HD\,111515 &$-0.19$ &$0.069$ &$-0.11$ &$0.202$ &$-0.12$ &$0.173$ \\
HD\,111777 &$-0.25$ &$0.005$ &$-0.30$ &$0.001$ &$-0.26$ &$0.003$ \\
HD\,11397 &$-0.02$ &$0.423$ &$0.37$ &$0.001$ &$0.41$ &$0.000$ \\
HD\,114853 &$0.16$ &$0.038$ &$-0.13$ &$0.068$ &$-0.20$ &$0.013$ \\
HD\,117207 &$-0.32$ &$0.002$ &$0.10$ &$0.198$ &$0.20$ &$0.038$ \\
HD\,119173 &$-0.22$ &$0.014$ &$0.03$ &$0.397$ &$0.12$ &$0.130$ \\
HD\,119949 &$0.05$ &$0.333$ &$0.08$ &$0.238$ &$0.03$ &$0.393$ \\
HD\,125072 &$0.58$ &$0.000$ &$-0.72$ &$0.000$ &$-0.86$ &$0.000$ \\
HD\,125595 &$0.70$ &$0.000$ &$0.34$ &$0.000$ &$0.08$ &$0.197$ \\
HD\,126525 &$-0.13$ &$0.064$ &$-0.13$ &$0.064$ &$-0.11$ &$0.106$ \\
HD\,126793 &$-0.23$ &$0.015$ &$0.03$ &$0.376$ &$0.10$ &$0.187$ \\
HD\,129642 &$0.75$ &$0.000$ &$-0.35$ &$0.000$ &$-0.70$ &$0.000$ \\
HD\,13060 &$0.80$ &$0.000$ &$0.36$ &$0.000$ &$-0.22$ &$0.018$ \\
HD\,133633 &$-0.37$ &$0.001$ &$-0.17$ &$0.086$ &$-0.02$ &$0.425$ \\
HD\,134060 &$-0.12$ &$0.063$ &$-0.33$ &$0.000$ &$-0.34$ &$0.000$ \\
HD\,134088 &$-0.36$ &$0.000$ &$-0.19$ &$0.037$ &$-0.04$ &$0.360$ \\
HD\,134606 &$-0.03$ &$0.306$ &$-0.04$ &$0.290$ &$-0.01$ &$0.421$ \\
HD\,136713 &$0.81$ &$0.000$ &$-0.26$ &$0.018$ &$-0.68$ &$0.000$ \\
HD\,137388 &$0.73$ &$0.000$ &$0.06$ &$0.274$ &$-0.37$ &$0.000$ \\
HD\,13808 &$0.91$ &$0.000$ &$0.75$ &$0.000$ &$-0.32$ &$0.000$ \\
HD\,1388 &$0.02$ &$0.384$ &$0.06$ &$0.251$ &$-0.01$ &$0.462$ \\
HD\,144628 &$0.91$ &$0.000$ &$0.64$ &$0.000$ &$-0.08$ &$0.104$ \\
HD\,145377 &$0.53$ &$0.000$ &$0.60$ &$0.000$ &$0.48$ &$0.000$ \\
HD\,1461 &$0.44$ &$0.000$ &$-0.16$ &$0.006$ &$-0.27$ &$0.000$ \\
HD\,148211 &$0.08$ &$0.246$ &$0.06$ &$0.320$ &$0.04$ &$0.384$ \\
HD\,149396 &$0.58$ &$0.000$ &$0.26$ &$0.004$ &$-0.14$ &$0.085$ \\
HD\,150433 &$-0.26$ &$0.003$ &$-0.43$ &$0.000$ &$-0.47$ &$0.000$ \\
HD\,15337 &$0.48$ &$0.000$ &$-0.26$ &$0.011$ &$-0.48$ &$0.000$ \\
HD\,154088 &$0.60$ &$0.000$ &$-0.35$ &$0.000$ &$-0.52$ &$0.000$ \\
HD\,154577 &$0.93$ &$0.000$ &$0.80$ &$0.000$ &$0.22$ &$0.000$ \\
HD\,157172 &$0.86$ &$0.000$ &$-0.34$ &$0.000$ &$-0.74$ &$0.000$ \\
HD\,157347 &$0.04$ &$0.359$ &$-0.05$ &$0.323$ &$-0.07$ &$0.263$ \\
HD\,157830 &$0.93$ &$0.000$ &$0.82$ &$0.000$ &$0.42$ &$0.001$ \\
HD\,16008 &$0.65$ &$0.000$ &$0.33$ &$0.003$ &$-0.04$ &$0.372$ \\
HD\,161098 &$0.74$ &$0.000$ &$0.25$ &$0.001$ &$-0.46$ &$0.000$ \\
HD\,172568 &$-0.27$ &$0.013$ &$-0.35$ &$0.002$ &$-0.31$ &$0.006$ \\
HD\,175607 &$0.09$ &$0.171$ &$0.24$ &$0.005$ &$0.19$ &$0.022$ \\
HD\,176986 &$0.52$ &$0.000$ &$0.00$ &$0.497$ &$-0.29$ &$0.000$ \\
HD\,177565 &$0.91$ &$0.000$ &$0.60$ &$0.000$ &$0.16$ &$0.028$ \\
HD\,17970 &$0.53$ &$0.000$ &$-0.02$ &$0.418$ &$-0.34$ &$0.002$ \\
HD\,181433 &$0.04$ &$0.287$ &$0.02$ &$0.394$ &$0.01$ &$0.467$ \\
HD\,181720 &$-0.07$ &$0.248$ &$-0.15$ &$0.076$ &$-0.06$ &$0.285$ \\
HD\,183658 &$0.16$ &$0.085$ &$-0.14$ &$0.121$ &$-0.19$ &$0.055$ \\
HD\,189567 &$0.65$ &$0.000$ &$0.40$ &$0.000$ &$0.13$ &$0.017$ \\
HD\,19467 &$-0.01$ &$0.461$ &$0.29$ &$0.007$ &$0.30$ &$0.007$ \\
HD\,197027 &$-0.26$ &$0.019$ &$-0.14$ &$0.134$ &$-0.04$ &$0.377$ \\
HD\,197197 &$-0.26$ &$0.010$ &$-0.40$ &$0.000$ &$-0.32$ &$0.002$ \\
HD\,199289 &$0.03$ &$0.408$ &$-0.08$ &$0.271$ &$-0.17$ &$0.082$ \\
HD\,199847 &$-0.31$ &$0.004$ &$-0.41$ &$0.000$ &$-0.33$ &$0.001$ \\
HD\,20003 &$0.71$ &$0.000$ &$-0.06$ &$0.201$ &$-0.63$ &$0.000$ \\
HD\,202206 &$0.89$ &$0.000$ &$0.53$ &$0.000$ &$0.10$ &$0.210$ \\
HD\,20407 &$0.05$ &$0.323$ &$0.06$ &$0.311$ &$0.04$ &$0.382$ \\
HD\,204313 &$0.54$ &$0.000$ &$-0.20$ &$0.022$ &$-0.41$ &$0.000$ \\
HD\,204941 &$0.89$ &$0.000$ &$0.74$ &$0.000$ &$-0.27$ &$0.004$ \\
HD\,206998 &$-0.66$ &$0.000$ &$-0.66$ &$0.000$ &$-0.58$ &$0.000$ \\
HD\,2071 &$0.85$ &$0.000$ &$0.58$ &$0.000$ &$0.01$ &$0.478$ \\
HD\,207129 &$0.89$ &$0.000$ &$0.29$ &$0.001$ &$-0.44$ &$0.000$ \\
HD\,20781 &$0.01$ &$0.461$ &$0.05$ &$0.207$ &$0.06$ &$0.174$ \\
HD\,20782 &$0.45$ &$0.000$ &$0.34$ &$0.001$ &$0.14$ &$0.103$ \\
HD\,207869 &$-0.29$ &$0.005$ &$-0.09$ &$0.203$ &$0.04$ &$0.375$ \\
HD\,208704 &$0.62$ &$0.000$ &$-0.01$ &$0.461$ &$-0.17$ &$0.078$ \\
HD\,210918 &$-0.23$ &$0.004$ &$-0.16$ &$0.028$ &$-0.14$ &$0.050$ \\
HD\,21132 &$0.14$ &$0.091$ &$-0.12$ &$0.123$ &$-0.15$ &$0.079$ \\
HD\,211415 &$0.54$ &$0.000$ &$-0.25$ &$0.022$ &$-0.43$ &$0.000$ \\
HD\,21209 &$0.79$ &$0.000$ &$0.38$ &$0.001$ &$-0.21$ &$0.051$ \\
HD\,215152 &$0.96$ &$0.000$ &$0.74$ &$0.000$ &$0.16$ &$0.002$ \\
HD\,21693 &$0.91$ &$0.000$ &$0.06$ &$0.182$ &$-0.66$ &$0.000$ \\
HD\,219828 &$0.04$ &$0.361$ &$-0.05$ &$0.319$ &$-0.09$ &$0.199$ \\
HD\,220507 &$-0.10$ &$0.090$ &$-0.08$ &$0.134$ &$-0.02$ &$0.377$ \\
HD\,222669 &$0.65$ &$0.000$ &$0.19$ &$0.061$ &$-0.01$ &$0.451$ \\
HD\,223171 &$0.43$ &$0.000$ &$-0.17$ &$0.024$ &$-0.38$ &$0.000$ \\
HD\,224817 &$-0.10$ &$0.151$ &$0.07$ &$0.243$ &$0.08$ &$0.211$ \\
HD\,22879 &$-0.31$ &$0.000$ &$-0.14$ &$0.052$ &$-0.09$ &$0.142$ \\
HD\,31103 &$0.42$ &$0.000$ &$0.47$ &$0.000$ &$0.38$ &$0.001$ \\
HD\,31128 &$-0.23$ &$0.001$ &$0.01$ &$0.437$ &$0.12$ &$0.041$ \\
HD\,31527 &$0.08$ &$0.105$ &$0.01$ &$0.446$ &$-0.02$ &$0.382$ \\
HD\,31822 &$0.34$ &$0.001$ &$0.24$ &$0.015$ &$0.12$ &$0.144$ \\
HD\,32564 &$-0.04$ &$0.290$ &$-0.02$ &$0.379$ &$-0.01$ &$0.452$ \\
HD\,35854 &$0.94$ &$0.000$ &$0.56$ &$0.000$ &$0.10$ &$0.180$ \\
HD\,36003 &$0.89$ &$0.000$ &$0.08$ &$0.202$ &$-0.58$ &$0.000$ \\
HD\,36379 &$0.18$ &$0.013$ &$-0.13$ &$0.064$ &$-0.21$ &$0.006$ \\
HD\,3823 &$-0.18$ &$0.016$ &$-0.13$ &$0.065$ &$-0.07$ &$0.196$ \\
HD\,38858 &$0.55$ &$0.000$ &$-0.01$ &$0.471$ &$-0.41$ &$0.000$ \\
HD\,39194 &$0.52$ &$0.000$ &$0.30$ &$0.000$ &$-0.04$ &$0.263$ \\
HD\,40307 &$0.93$ &$0.000$ &$0.42$ &$0.000$ &$-0.46$ &$0.000$ \\
HD\,40865 &$0.24$ &$0.022$ &$-0.42$ &$0.000$ &$-0.47$ &$0.000$ \\
HD\,41248 &$0.29$ &$0.000$ &$0.12$ &$0.065$ &$-0.05$ &$0.256$ \\
HD\,4308 &$0.34$ &$0.000$ &$0.01$ &$0.429$ &$-0.13$ &$0.057$ \\
HD\,45184 &$0.92$ &$0.000$ &$0.05$ &$0.251$ &$-0.54$ &$0.000$ \\
HD\,45289 &$0.26$ &$0.020$ &$0.25$ &$0.026$ &$0.15$ &$0.118$ \\
HD\,45364 &$0.39$ &$0.000$ &$-0.20$ &$0.015$ &$-0.34$ &$0.000$ \\
HD\,47186 &$-0.14$ &$0.044$ &$-0.13$ &$0.065$ &$-0.10$ &$0.106$ \\
HD\,51608 &$0.60$ &$0.000$ &$-0.09$ &$0.093$ &$-0.33$ &$0.000$ \\
HD\,5388 &$-0.10$ &$0.196$ &$-0.03$ &$0.402$ &$-0.00$ &$0.486$ \\
HD\,56274 &$0.62$ &$0.000$ &$0.71$ &$0.000$ &$0.40$ &$0.000$ \\
HD\,564 &$0.33$ &$0.000$ &$0.14$ &$0.075$ &$-0.06$ &$0.277$ \\
HD\,59468 &$0.37$ &$0.000$ &$-0.27$ &$0.000$ &$-0.37$ &$0.000$ \\
HD\,59711 &$0.61$ &$0.000$ &$0.28$ &$0.013$ &$0.14$ &$0.129$ \\
HD\,61986 &$0.67$ &$0.000$ &$0.29$ &$0.003$ &$-0.43$ &$0.000$ \\
HD\,63765 &$0.90$ &$0.000$ &$0.76$ &$0.000$ &$0.38$ &$0.001$ \\
HD\,65277 &$0.86$ &$0.000$ &$0.39$ &$0.001$ &$-0.35$ &$0.002$ \\
HD\,65907 &$0.01$ &$0.410$ &$-0.25$ &$0.000$ &$-0.25$ &$0.000$ \\
HD\,68978 &$0.79$ &$0.000$ &$0.41$ &$0.000$ &$-0.27$ &$0.002$ \\
HD\,69611 &$-0.18$ &$0.067$ &$-0.33$ &$0.003$ &$-0.32$ &$0.004$ \\
HD\,69830 &$0.56$ &$0.000$ &$-0.09$ &$0.068$ &$-0.49$ &$0.000$ \\
HD\,71334 &$0.18$ &$0.079$ &$0.10$ &$0.228$ &$0.07$ &$0.285$ \\
HD\,71835 &$0.93$ &$0.000$ &$0.13$ &$0.052$ &$-0.67$ &$0.000$ \\
HD\,7199 &$0.67$ &$0.000$ &$-0.80$ &$0.000$ &$-0.85$ &$0.000$ \\
HD\,72673 &$0.93$ &$0.000$ &$0.52$ &$0.000$ &$-0.25$ &$0.001$ \\
HD\,73524 &$0.22$ &$0.003$ &$-0.44$ &$0.000$ &$-0.46$ &$0.000$ \\
HD\,7449 &$0.77$ &$0.000$ &$0.46$ &$0.000$ &$-0.14$ &$0.064$ \\
HD\,77110 &$-0.20$ &$0.020$ &$-0.22$ &$0.013$ &$-0.20$ &$0.024$ \\
HD\,78429 &$0.51$ &$0.000$ &$-0.09$ &$0.156$ &$-0.48$ &$0.000$ \\
HD\,79601 &$0.24$ &$0.016$ &$-0.03$ &$0.401$ &$-0.09$ &$0.201$ \\
HD\,82342 &$0.55$ &$0.000$ &$-0.26$ &$0.018$ &$-0.51$ &$0.000$ \\
HD\,82516 &$0.80$ &$0.000$ &$-0.47$ &$0.000$ &$-0.67$ &$0.000$ \\
HD\,82943 &$0.74$ &$0.000$ &$-0.54$ &$0.000$ &$-0.67$ &$0.000$ \\
HD\,85390 &$0.72$ &$0.000$ &$0.19$ &$0.023$ &$-0.19$ &$0.022$ \\
HD\,87838 &$-0.19$ &$0.009$ &$0.01$ &$0.445$ &$0.07$ &$0.205$ \\
HD\,8828 &$-0.30$ &$0.004$ &$-0.16$ &$0.081$ &$-0.08$ &$0.251$ \\
HD\,88725 &$0.10$ &$0.161$ &$-0.15$ &$0.080$ &$-0.15$ &$0.080$ \\
HD\,89839 &$0.04$ &$0.355$ &$0.07$ &$0.263$ &$0.04$ &$0.352$ \\
HD\,90156 &$-0.02$ &$0.396$ &$0.02$ &$0.416$ &$0.01$ &$0.446$ \\
HD\,92719 &$0.91$ &$0.000$ &$0.72$ &$0.000$ &$0.26$ &$0.016$ \\
HD\,93083 &$0.63$ &$0.000$ &$-0.27$ &$0.001$ &$-0.54$ &$0.000$ \\
HD\,93385 &$-0.07$ &$0.153$ &$-0.07$ &$0.155$ &$-0.03$ &$0.349$ \\
HD\,95456 &$0.69$ &$0.000$ &$-0.67$ &$0.000$ &$-0.78$ &$0.000$ \\
HD\,9578 &$0.52$ &$0.000$ &$0.45$ &$0.000$ &$0.34$ &$0.002$ \\
HD\,96423 &$0.15$ &$0.106$ &$0.11$ &$0.172$ &$0.07$ &$0.281$ \\
HD\,967 &$0.01$ &$0.462$ &$-0.08$ &$0.252$ &$-0.10$ &$0.202$ \\
HD\,96700 &$0.09$ &$0.085$ &$-0.14$ &$0.018$ &$-0.17$ &$0.007$ \\
HD\,97037 &$0.08$ &$0.255$ &$0.01$ &$0.457$ &$-0.01$ &$0.479$ \\
HD\,97343 &$0.08$ &$0.252$ &$-0.05$ &$0.330$ &$-0.03$ &$0.390$ \\
\end{longtable}
\tablefoot{
$\rho^{06}$, $\rho^{16}$, and $\rho^{W}$ are the correlation coefficients between \sca{} and \nsha{}, \wsha{}, and $S_{\text{H}\alpha W}$, respectively.$p$-value is the probability of having an equal or higher correlation coefficient (see text). $p$-values of $0.000$ means that the value is lower than $10^{-3}$.
}
}

\longtab[3]{
\tiny
\begin{longtable}{lccccccccccccc}

\caption{\label{tab:corr_table_short}Spearman correlation coefficients for the short-term dataset.} \\

\hline\hline
\multicolumn{1}{l}{Star}&
\multicolumn{1}{c}{$\langle S_\text{CaII} \rangle_\text{max}$}&
\multicolumn{1}{c}{$\langle S_\text{CaII} \rangle_\text{min}$}&
\multicolumn{1}{c}{$\sigma(S_\text{CaII})_\text{max}$}&
\multicolumn{1}{c}{$\sigma(S_\text{CaII})_\text{min}$}&
\multicolumn{1}{c}{$\rho^{06}_\mathrm{max}$}&
\multicolumn{1}{c}{$p$-value$^{06}_\mathrm{max}$}&
\multicolumn{1}{c}{$\rho^{06}_\mathrm{min}$}&
\multicolumn{1}{c}{$p$-value$^{06}_\mathrm{min}$}&
\multicolumn{1}{c}{$\rho^{16}_\mathrm{max}$}&
\multicolumn{1}{c}{$p$-value$^{16}_\mathrm{max}$}&
\multicolumn{1}{c}{$\rho^{16}_\mathrm{min}$}&
\multicolumn{1}{c}{$p$-value$^{16}_\mathrm{min}$}
\\

\hline
\endfirsthead
\caption{continued.} \\
\hline\hline
\multicolumn{1}{l}{Star}&
\multicolumn{1}{c}{$\langle S_\text{CaII} \rangle_\text{max}$}&
\multicolumn{1}{c}{$\langle S_\text{CaII} \rangle_\text{min}$}&
\multicolumn{1}{c}{$\sigma(S_\text{CaII})_\text{max}$}&
\multicolumn{1}{c}{$\sigma(S_\text{CaII})_\text{min}$}&
\multicolumn{1}{c}{$\rho^{06}_\mathrm{max}$}&
\multicolumn{1}{c}{$p$-value$^{06}_\mathrm{max}$}&
\multicolumn{1}{c}{$\rho^{06}_\mathrm{min}$}&
\multicolumn{1}{c}{$p$-value$^{06}_\mathrm{min}$}&
\multicolumn{1}{c}{$\rho^{16}_\mathrm{max}$}&
\multicolumn{1}{c}{$p$-value$^{16}_\mathrm{max}$}&
\multicolumn{1}{c}{$\rho^{16}_\mathrm{min}$}&
\multicolumn{1}{c}{$p$-value$^{16}_\mathrm{min}$}
\\

\hline
\endhead
\hline
\endfoot
18\,Pup &$0.1255$ &$0.1248$ &$0.0004$ &$0.0004$ &$-0.34$ &$0.087$ &$0.27$ &$0.183$ &$-0.20$ &$0.215$ &$0.66$ &$0.013$ \\
nu.02\,Lup &$0.1377$ &$0.1360$ &$0.0004$ &$0.0010$ &$-0.21$ &$0.256$ &$-0.46$ &$0.002$ &$0.14$ &$0.334$ &$0.12$ &$0.223$ \\
omi02\,Eri &$0.1615$ &$0.1421$ &$0.0063$ &$0.0063$ &$0.92$ &$0.000$ &$0.54$ &$0.029$ &$0.59$ &$0.010$ &$0.21$ &$0.235$ \\
zet02\,Ret &$0.1461$ &$0.1402$ &$0.0017$ &$0.0012$ &$0.32$ &$0.186$ &$0.43$ &$0.147$ &$-0.27$ &$0.224$ &$0.21$ &$0.298$ \\
BD-08\,2823 &$0.4608$ &$0.3858$ &$0.0257$ &$0.0165$ &$-0.03$ &$0.439$ &$0.08$ &$0.340$ &$0.12$ &$0.286$ &$0.10$ &$0.311$ \\
CoRoT-7 &$0.3072$ &$0.2383$ &$0.0208$ &$0.0094$ &$0.51$ &$0.000$ &$0.50$ &$0.007$ &$0.45$ &$0.000$ &$0.36$ &$0.038$ \\
HD\,10180 &$0.1288$ &$0.1257$ &$0.0009$ &$0.0011$ &$-0.12$ &$0.208$ &$0.22$ &$0.250$ &$-0.01$ &$0.474$ &$0.09$ &$0.392$ \\
HD\,101930 &$0.1580$ &$0.1482$ &$0.0037$ &$0.0061$ &$-0.12$ &$0.371$ &$0.48$ &$0.076$ &$0.27$ &$0.224$ &$0.18$ &$0.298$ \\
HD\,102117 &$0.1168$ &$0.1168$ &$0.0006$ &$0.0006$ &$0.68$ &$0.027$ &$0.68$ &$0.027$ &$-0.08$ &$0.407$ &$-0.08$ &$0.406$ \\
HD\,102365 &$0.1395$ &$0.1376$ &$0.0010$ &$0.0003$ &$0.60$ &$0.011$ &$0.10$ &$0.389$ &$-0.18$ &$0.241$ &$-0.10$ &$0.388$ \\
HD\,102438 &$0.1392$ &$0.1359$ &$0.0005$ &$0.0005$ &$0.26$ &$0.243$ &$0.64$ &$0.057$ &$0.19$ &$0.308$ &$0.00$ &$0.500$ \\
HD\,103774 &$0.1492$ &$0.1302$ &$0.0091$ &$0.0061$ &$-0.03$ &$0.454$ &$0.48$ &$0.054$ &$0.59$ &$0.020$ &$0.32$ &$0.142$ \\
HD\,104067 &$0.3071$ &$0.2876$ &$0.0194$ &$0.0162$ &$0.90$ &$0.008$ &$0.67$ &$0.002$ &$0.33$ &$0.187$ &$0.42$ &$0.032$ \\
HD\,104800 &$0.1408$ &$0.1352$ &$0.0019$ &$0.0030$ &$0.20$ &$0.287$ &$-0.17$ &$0.206$ &$0.17$ &$0.320$ &$0.05$ &$0.399$ \\
HD\,106116 &$0.1312$ &$0.1228$ &$0.0016$ &$0.0010$ &$0.70$ &$0.024$ &$-0.04$ &$0.439$ &$0.85$ &$0.008$ &$0.47$ &$0.045$ \\
HD\,107094 &$0.1603$ &$0.1581$ &$0.0067$ &$0.0032$ &$-0.01$ &$0.489$ &$0.16$ &$0.220$ &$-0.07$ &$0.395$ &$0.27$ &$0.105$ \\
HD\,109200 &$0.1665$ &$0.1417$ &$0.0036$ &$0.0003$ &$0.79$ &$0.000$ &$-0.23$ &$0.254$ &$0.31$ &$0.021$ &$-0.17$ &$0.319$ \\
HD\,109271 &$0.1391$ &$0.1255$ &$0.0092$ &$0.0068$ &$0.35$ &$0.075$ &$0.12$ &$0.235$ &$0.30$ &$0.108$ &$-0.21$ &$0.099$ \\
HD\,111515 &$0.1430$ &$0.1388$ &$0.0006$ &$0.0013$ &$-0.02$ &$0.481$ &$-0.09$ &$0.359$ &$0.47$ &$0.091$ &$-0.07$ &$0.389$ \\
HD\,111777 &$0.1374$ &$0.1365$ &$0.0020$ &$0.0025$ &$-0.20$ &$0.052$ &$0.10$ &$0.400$ &$-0.31$ &$0.007$ &$-0.69$ &$0.034$ \\
HD\,11397 &$0.1519$ &$0.1468$ &$0.0016$ &$0.0077$ &$0.49$ &$0.020$ &$-0.07$ &$0.378$ &$0.35$ &$0.072$ &$0.22$ &$0.170$ \\
HD\,114853 &$0.1398$ &$0.1350$ &$0.0015$ &$0.0013$ &$-0.50$ &$0.109$ &$0.18$ &$0.114$ &$-0.14$ &$0.363$ &$0.04$ &$0.386$ \\
HD\,117207 &$0.1237$ &$0.1202$ &$0.0005$ &$0.0015$ &$-0.39$ &$0.107$ &$-0.61$ &$0.069$ &$-0.05$ &$0.443$ &$0.46$ &$0.128$ \\
HD\,119173 &$0.1413$ &$0.1380$ &$0.0008$ &$0.0044$ &$0.43$ &$0.147$ &$0.03$ &$0.446$ &$0.32$ &$0.214$ &$0.18$ &$0.186$ \\
HD\,119949 &$0.1256$ &$0.1232$ &$0.0005$ &$0.0013$ &$-0.86$ &$0.018$ &$-0.54$ &$0.092$ &$-0.07$ &$0.431$ &$-0.14$ &$0.362$ \\
HD\,125072 &$0.2464$ &$0.1862$ &$0.0138$ &$0.0285$ &$0.69$ &$0.004$ &$-0.54$ &$0.094$ &$-0.37$ &$0.087$ &$-0.86$ &$0.018$ \\
HD\,125595 &$0.4329$ &$0.3757$ &$0.0286$ &$0.0178$ &$0.48$ &$0.024$ &$0.52$ &$0.032$ &$-0.37$ &$0.062$ &$0.42$ &$0.069$ \\
HD\,126525 &$0.1338$ &$0.1297$ &$0.0011$ &$0.0013$ &$0.33$ &$0.159$ &$0.68$ &$0.049$ &$0.20$ &$0.275$ &$0.14$ &$0.364$ \\
HD\,126793 &$0.1310$ &$0.1282$ &$0.0009$ &$0.0017$ &$0.16$ &$0.300$ &$-0.90$ &$0.008$ &$0.10$ &$0.375$ &$-0.71$ &$0.031$ \\
HD\,129642 &$0.1784$ &$0.1316$ &$0.0029$ &$0.0063$ &$0.37$ &$0.152$ &$-0.19$ &$0.240$ &$0.47$ &$0.093$ &$-0.62$ &$0.010$ \\
HD\,13060 &$0.1880$ &$0.1607$ &$0.0134$ &$0.0066$ &$0.60$ &$0.000$ &$0.03$ &$0.462$ &$0.11$ &$0.240$ &$-0.18$ &$0.302$ \\
HD\,133633 &$0.1375$ &$0.1365$ &$0.0028$ &$0.0020$ &$-0.24$ &$0.105$ &$0.07$ &$0.388$ &$-0.31$ &$0.054$ &$-0.32$ &$0.093$ \\
HD\,134060 &$0.1283$ &$0.1246$ &$0.0010$ &$0.0006$ &$-0.32$ &$0.157$ &$-0.68$ &$0.019$ &$0.38$ &$0.115$ &$-0.30$ &$0.186$ \\
HD\,134088 &$0.1416$ &$0.1378$ &$0.0006$ &$0.0023$ &$0.20$ &$0.287$ &$-0.76$ &$0.012$ &$-0.38$ &$0.139$ &$-0.66$ &$0.024$ \\
HD\,134606 &$0.1211$ &$0.1191$ &$0.0007$ &$0.0008$ &$0.24$ &$0.111$ &$0.14$ &$0.338$ &$-0.25$ &$0.104$ &$0.25$ &$0.230$ \\
HD\,136713 &$0.2841$ &$0.2744$ &$0.0164$ &$0.0145$ &$0.87$ &$0.001$ &$0.77$ &$0.010$ &$0.54$ &$0.030$ &$0.24$ &$0.240$ \\
HD\,137388 &$0.1972$ &$0.1559$ &$0.0117$ &$0.0086$ &$0.38$ &$0.101$ &$0.39$ &$0.035$ &$-0.27$ &$0.181$ &$0.62$ &$0.002$ \\
HD\,13808 &$0.2228$ &$0.1555$ &$0.0088$ &$0.0037$ &$0.73$ &$0.001$ &$0.36$ &$0.003$ &$0.72$ &$0.001$ &$-0.16$ &$0.114$ \\
HD\,1388 &$0.1291$ &$0.1259$ &$0.0006$ &$0.0005$ &$0.49$ &$0.011$ &$0.08$ &$0.361$ &$0.28$ &$0.096$ &$0.06$ &$0.391$ \\
HD\,144628 &$0.1738$ &$0.1495$ &$0.0071$ &$0.0021$ &$0.76$ &$0.000$ &$0.60$ &$0.047$ &$0.85$ &$0.000$ &$0.12$ &$0.372$ \\
HD\,145377 &$0.2139$ &$0.1948$ &$0.0075$ &$0.0090$ &$0.67$ &$0.030$ &$0.75$ &$0.002$ &$0.83$ &$0.010$ &$0.56$ &$0.018$ \\
HD\,1461 &$0.1293$ &$0.1259$ &$0.0010$ &$0.0008$ &$-0.32$ &$0.135$ &$-0.01$ &$0.486$ &$0.02$ &$0.470$ &$0.20$ &$0.138$ \\
HD\,148211 &$0.1302$ &$0.1280$ &$0.0016$ &$0.0023$ &$-0.18$ &$0.300$ &$0.24$ &$0.124$ &$-0.32$ &$0.186$ &$0.19$ &$0.179$ \\
HD\,149396 &$0.2230$ &$0.1913$ &$0.0127$ &$0.0159$ &$0.58$ &$0.041$ &$0.28$ &$0.056$ &$0.36$ &$0.142$ &$0.10$ &$0.284$ \\
HD\,150433 &$0.1348$ &$0.1310$ &$0.0004$ &$0.0005$ &$0.60$ &$0.034$ &$0.35$ &$0.150$ &$0.18$ &$0.297$ &$0.16$ &$0.311$ \\
HD\,15337 &$0.1697$ &$0.1594$ &$0.0047$ &$0.0052$ &$0.57$ &$0.053$ &$0.82$ &$0.022$ &$0.48$ &$0.089$ &$0.50$ &$0.111$ \\
HD\,154088 &$0.1369$ &$0.1235$ &$0.0048$ &$0.0007$ &$0.83$ &$0.010$ &$0.15$ &$0.296$ &$0.42$ &$0.120$ &$0.26$ &$0.176$ \\
HD\,154577 &$0.1995$ &$0.1710$ &$0.0048$ &$0.0034$ &$0.73$ &$0.000$ &$0.91$ &$0.000$ &$0.82$ &$0.000$ &$0.75$ &$0.001$ \\
HD\,157172 &$0.1524$ &$0.1318$ &$0.0021$ &$0.0015$ &$0.38$ &$0.128$ &$0.07$ &$0.364$ &$-0.08$ &$0.407$ &$-0.20$ &$0.164$ \\
HD\,157347 &$0.1297$ &$0.1266$ &$0.0004$ &$0.0011$ &$-0.10$ &$0.380$ &$0.04$ &$0.449$ &$-0.70$ &$0.018$ &$-0.22$ &$0.253$ \\
HD\,157830 &$0.1762$ &$0.1634$ &$0.0045$ &$0.0042$ &$0.55$ &$0.043$ &$0.86$ &$0.018$ &$0.48$ &$0.065$ &$0.86$ &$0.018$ \\
HD\,16008 &$0.1578$ &$0.1434$ &$0.0035$ &$0.0087$ &$0.81$ &$0.003$ &$-0.12$ &$0.376$ &$0.65$ &$0.013$ &$-0.86$ &$0.011$ \\
HD\,161098 &$0.1506$ &$0.1391$ &$0.0017$ &$0.0023$ &$0.51$ &$0.009$ &$0.10$ &$0.334$ &$0.47$ &$0.014$ &$0.08$ &$0.367$ \\
HD\,172568 &$0.1434$ &$0.1379$ &$0.0011$ &$0.0050$ &$-0.38$ &$0.128$ &$-0.03$ &$0.440$ &$-0.42$ &$0.105$ &$-0.06$ &$0.373$ \\
HD\,175607 &$0.1482$ &$0.1421$ &$0.0019$ &$0.0046$ &$0.65$ &$0.026$ &$-0.55$ &$0.075$ &$0.65$ &$0.027$ &$-0.07$ &$0.425$ \\
HD\,176986 &$0.2298$ &$0.2072$ &$0.0098$ &$0.0087$ &$0.54$ &$0.007$ &$0.21$ &$0.130$ &$0.44$ &$0.022$ &$-0.05$ &$0.402$ \\
HD\,177565 &$0.1577$ &$0.1371$ &$0.0024$ &$0.0038$ &$0.82$ &$0.000$ &$-0.21$ &$0.200$ &$0.45$ &$0.009$ &$-0.30$ &$0.108$ \\
HD\,17970 &$0.1470$ &$0.1396$ &$0.0039$ &$0.0023$ &$0.86$ &$0.017$ &$0.25$ &$0.208$ &$0.25$ &$0.269$ &$-0.32$ &$0.157$ \\
HD\,181433 &$0.1371$ &$0.1325$ &$0.0037$ &$0.0046$ &$-0.56$ &$0.016$ &$0.29$ &$0.245$ &$0.30$ &$0.133$ &$0.11$ &$0.396$ \\
HD\,181720 &$0.1274$ &$0.1226$ &$0.0010$ &$0.0012$ &$-0.31$ &$0.205$ &$0.11$ &$0.398$ &$-0.43$ &$0.130$ &$-0.82$ &$0.022$ \\
HD\,183658 &$0.1336$ &$0.1296$ &$0.0015$ &$0.0017$ &$-0.21$ &$0.299$ &$0.75$ &$0.033$ &$-0.21$ &$0.301$ &$0.54$ &$0.097$ \\
HD\,189567 &$0.1463$ &$0.1403$ &$0.0033$ &$0.0018$ &$0.74$ &$0.004$ &$0.96$ &$0.010$ &$0.63$ &$0.012$ &$0.29$ &$0.241$ \\
HD\,19467 &$0.1292$ &$0.1280$ &$0.0015$ &$0.0001$ &$-0.15$ &$0.335$ &$-0.64$ &$0.057$ &$-0.15$ &$0.335$ &$0.32$ &$0.215$ \\
HD\,197027 &$0.1297$ &$0.1265$ &$0.0029$ &$0.0010$ &$-0.34$ &$0.068$ &$-0.14$ &$0.352$ &$-0.48$ &$0.018$ &$0.02$ &$0.475$ \\
HD\,197197 &$0.1232$ &$0.1185$ &$0.0010$ &$0.0012$ &$0.35$ &$0.130$ &$-0.08$ &$0.362$ &$0.15$ &$0.311$ &$-0.26$ &$0.123$ \\
HD\,199289 &$0.1344$ &$0.1337$ &$0.0011$ &$0.0028$ &$-0.36$ &$0.172$ &$-0.19$ &$0.164$ &$-0.74$ &$0.023$ &$-0.02$ &$0.465$ \\
HD\,199847 &$0.1240$ &$0.1214$ &$0.0031$ &$0.0054$ &$-0.64$ &$0.002$ &$0.31$ &$0.112$ &$-0.81$ &$0.000$ &$0.13$ &$0.310$ \\
HD\,20003 &$0.1597$ &$0.1315$ &$0.0034$ &$0.0023$ &$0.86$ &$0.018$ &$0.18$ &$0.168$ &$0.68$ &$0.049$ &$0.05$ &$0.395$ \\
HD\,202206 &$0.2005$ &$0.1801$ &$0.0066$ &$0.0054$ &$0.67$ &$0.039$ &$0.61$ &$0.002$ &$0.48$ &$0.105$ &$-0.30$ &$0.075$ \\
HD\,20407 &$0.1396$ &$0.1376$ &$0.0009$ &$0.0013$ &$-0.03$ &$0.454$ &$0.38$ &$0.100$ &$0.50$ &$0.050$ &$-0.01$ &$0.491$ \\
HD\,204313 &$0.1313$ &$0.1272$ &$0.0018$ &$0.0014$ &$0.28$ &$0.129$ &$0.31$ &$0.207$ &$0.30$ &$0.115$ &$0.31$ &$0.207$ \\
HD\,204941 &$0.1803$ &$0.1522$ &$0.0053$ &$0.0038$ &$0.33$ &$0.160$ &$0.23$ &$0.254$ &$-0.14$ &$0.337$ &$-0.12$ &$0.370$ \\
HD\,206998 &$0.1288$ &$0.1209$ &$0.0024$ &$0.0049$ &$0.36$ &$0.191$ &$-0.15$ &$0.314$ &$0.46$ &$0.130$ &$-0.39$ &$0.106$ \\
HD\,2071 &$0.1541$ &$0.1408$ &$0.0033$ &$0.0012$ &$0.93$ &$0.012$ &$0.57$ &$0.081$ &$0.89$ &$0.015$ &$0.36$ &$0.192$ \\
HD\,207129 &$0.1475$ &$0.1347$ &$0.0026$ &$0.0029$ &$0.92$ &$0.002$ &$0.41$ &$0.059$ &$0.66$ &$0.019$ &$0.03$ &$0.459$ \\
HD\,20781 &$0.1332$ &$0.1302$ &$0.0009$ &$0.0028$ &$0.07$ &$0.425$ &$-0.22$ &$0.162$ &$-0.20$ &$0.287$ &$-0.16$ &$0.233$ \\
HD\,20782 &$0.1421$ &$0.1400$ &$0.0016$ &$0.0012$ &$0.29$ &$0.243$ &$0.75$ &$0.003$ &$0.36$ &$0.188$ &$0.58$ &$0.018$ \\
HD\,207869 &$0.1413$ &$0.1360$ &$0.0017$ &$0.0037$ &$0.29$ &$0.223$ &$-0.38$ &$0.067$ &$0.02$ &$0.475$ &$-0.30$ &$0.114$ \\
HD\,208704 &$0.1422$ &$0.1345$ &$0.0021$ &$0.0014$ &$0.50$ &$0.092$ &$0.85$ &$0.005$ &$-0.43$ &$0.129$ &$0.02$ &$0.478$ \\
HD\,210918 &$0.1286$ &$0.1271$ &$0.0015$ &$0.0007$ &$-0.86$ &$0.017$ &$-0.27$ &$0.152$ &$-0.36$ &$0.191$ &$0.50$ &$0.030$ \\
HD\,21132 &$0.1361$ &$0.1313$ &$0.0045$ &$0.0012$ &$0.16$ &$0.283$ &$-0.25$ &$0.136$ &$-0.08$ &$0.394$ &$-0.17$ &$0.237$ \\
HD\,211415 &$0.1351$ &$0.1349$ &$0.0005$ &$0.0007$ &$0.13$ &$0.351$ &$0.29$ &$0.243$ &$-0.05$ &$0.443$ &$0.54$ &$0.093$ \\
HD\,21209 &$0.3427$ &$0.3223$ &$0.0129$ &$0.0100$ &$0.90$ &$0.001$ &$0.82$ &$0.023$ &$0.42$ &$0.065$ &$0.07$ &$0.432$ \\
HD\,215152 &$0.2452$ &$0.1848$ &$0.0118$ &$0.0040$ &$0.95$ &$0.000$ &$0.62$ &$0.002$ &$0.90$ &$0.000$ &$0.46$ &$0.017$ \\
HD\,21693 &$0.1640$ &$0.1434$ &$0.0037$ &$0.0027$ &$0.64$ &$0.000$ &$0.39$ &$0.015$ &$0.09$ &$0.285$ &$0.05$ &$0.401$ \\
HD\,219828 &$0.1149$ &$0.1136$ &$0.0010$ &$0.0033$ &$-0.06$ &$0.404$ &$0.83$ &$0.009$ &$-0.06$ &$0.392$ &$0.05$ &$0.444$ \\
HD\,220507 &$0.1242$ &$0.1213$ &$0.0008$ &$0.0011$ &$0.13$ &$0.275$ &$0.35$ &$0.019$ &$0.46$ &$0.015$ &$0.26$ &$0.065$ \\
HD\,222669 &$0.1485$ &$0.1440$ &$0.0030$ &$0.0013$ &$0.68$ &$0.049$ &$0.71$ &$0.016$ &$0.04$ &$0.465$ &$-0.15$ &$0.325$ \\
HD\,223171 &$0.1294$ &$0.1241$ &$0.0026$ &$0.0015$ &$-0.30$ &$0.200$ &$0.70$ &$0.013$ &$-0.23$ &$0.253$ &$0.05$ &$0.432$ \\
HD\,224817 &$0.1287$ &$0.1267$ &$0.0014$ &$0.0024$ &$0.21$ &$0.228$ &$0.33$ &$0.071$ &$0.44$ &$0.064$ &$0.53$ &$0.010$ \\
HD\,22879 &$0.1354$ &$0.1317$ &$0.0003$ &$0.0006$ &$0.07$ &$0.421$ &$0.07$ &$0.390$ &$0.36$ &$0.141$ &$0.24$ &$0.164$ \\
HD\,31103 &$0.1894$ &$0.1831$ &$0.0127$ &$0.0138$ &$0.51$ &$0.054$ &$0.50$ &$0.111$ &$0.70$ &$0.013$ &$0.21$ &$0.300$ \\
HD\,31128 &$0.1347$ &$0.1308$ &$0.0010$ &$0.0026$ &$0.01$ &$0.493$ &$0.10$ &$0.284$ &$-0.19$ &$0.285$ &$-0.26$ &$0.070$ \\
HD\,31527 &$0.1335$ &$0.1307$ &$0.0009$ &$0.0004$ &$-0.15$ &$0.303$ &$0.64$ &$0.045$ &$0.05$ &$0.424$ &$0.55$ &$0.072$ \\
HD\,31822 &$0.1446$ &$0.1428$ &$0.0011$ &$0.0010$ &$0.17$ &$0.221$ &$0.64$ &$0.056$ &$0.30$ &$0.093$ &$0.46$ &$0.128$ \\
HD\,32564 &$0.1326$ &$0.1304$ &$0.0017$ &$0.0029$ &$-0.13$ &$0.250$ &$-0.23$ &$0.211$ &$-0.13$ &$0.249$ &$0.11$ &$0.351$ \\
HD\,35854 &$0.2501$ &$0.2174$ &$0.0084$ &$0.0048$ &$0.65$ &$0.020$ &$0.89$ &$0.013$ &$0.52$ &$0.050$ &$0.82$ &$0.023$ \\
HD\,36003 &$0.3687$ &$0.2747$ &$0.0218$ &$0.0082$ &$0.90$ &$0.003$ &$-0.11$ &$0.317$ &$0.81$ &$0.008$ &$-0.35$ &$0.056$ \\
HD\,36379 &$0.1263$ &$0.1226$ &$0.0005$ &$0.0007$ &$0.10$ &$0.379$ &$0.36$ &$0.192$ &$-0.42$ &$0.103$ &$-0.07$ &$0.431$ \\
HD\,3823 &$0.1248$ &$0.1222$ &$0.0004$ &$0.0006$ &$0.32$ &$0.105$ &$0.13$ &$0.315$ &$-0.19$ &$0.226$ &$0.68$ &$0.006$ \\
HD\,38858 &$0.1463$ &$0.1404$ &$0.0014$ &$0.0008$ &$0.80$ &$0.003$ &$0.48$ &$0.104$ &$-0.42$ &$0.072$ &$0.19$ &$0.307$ \\
HD\,39194 &$0.1466$ &$0.1412$ &$0.0035$ &$0.0009$ &$0.67$ &$0.000$ &$-0.27$ &$0.176$ &$0.43$ &$0.002$ &$0.13$ &$0.323$ \\
HD\,40307 &$0.2163$ &$0.1568$ &$0.0066$ &$0.0013$ &$0.89$ &$0.000$ &$0.36$ &$0.194$ &$0.45$ &$0.026$ &$0.71$ &$0.040$ \\
HD\,40865 &$0.1411$ &$0.1357$ &$0.0012$ &$0.0021$ &$0.68$ &$0.021$ &$-0.15$ &$0.252$ &$0.01$ &$0.493$ &$0.28$ &$0.102$ \\
HD\,41248 &$0.1452$ &$0.1421$ &$0.0020$ &$0.0042$ &$0.32$ &$0.184$ &$0.18$ &$0.206$ &$-0.07$ &$0.425$ &$0.19$ &$0.200$ \\
HD\,4308 &$0.1377$ &$0.1347$ &$0.0015$ &$0.0019$ &$-0.06$ &$0.410$ &$0.15$ &$0.257$ &$0.10$ &$0.358$ &$-0.02$ &$0.464$ \\
HD\,45184 &$0.1473$ &$0.1349$ &$0.0020$ &$0.0006$ &$0.87$ &$0.005$ &$0.22$ &$0.249$ &$0.77$ &$0.010$ &$0.24$ &$0.237$ \\
HD\,45289 &$0.1254$ &$0.1252$ &$0.0004$ &$0.0006$ &$0.32$ &$0.216$ &$-0.10$ &$0.400$ &$0.57$ &$0.081$ &$-0.07$ &$0.426$ \\
HD\,45364 &$0.1444$ &$0.1392$ &$0.0072$ &$0.0020$ &$0.77$ &$0.014$ &$0.55$ &$0.041$ &$0.00$ &$0.500$ &$-0.18$ &$0.282$ \\
HD\,47186 &$0.1253$ &$0.1229$ &$0.0003$ &$0.0014$ &$0.38$ &$0.128$ &$-0.18$ &$0.330$ &$-0.42$ &$0.104$ &$-0.71$ &$0.041$ \\
HD\,51608 &$0.1441$ &$0.1311$ &$0.0025$ &$0.0020$ &$0.55$ &$0.073$ &$-0.47$ &$0.094$ &$0.76$ &$0.021$ &$0.07$ &$0.425$ \\
HD\,5388 &$0.1194$ &$0.1174$ &$0.0054$ &$0.0076$ &$-0.00$ &$0.499$ &$-0.20$ &$0.165$ &$-0.26$ &$0.120$ &$-0.01$ &$0.483$ \\
HD\,56274 &$0.1530$ &$0.1468$ &$0.0016$ &$0.0009$ &$0.33$ &$0.028$ &$0.59$ &$0.038$ &$0.67$ &$0.000$ &$-0.14$ &$0.339$ \\
HD\,564 &$0.1535$ &$0.1351$ &$0.0113$ &$0.0106$ &$-0.18$ &$0.254$ &$-0.20$ &$0.196$ &$-0.12$ &$0.325$ &$-0.34$ &$0.069$ \\
HD\,59468 &$0.1335$ &$0.1307$ &$0.0013$ &$0.0015$ &$0.75$ &$0.013$ &$0.46$ &$0.004$ &$0.94$ &$0.003$ &$-0.57$ &$0.000$ \\
HD\,59711 &$0.1382$ &$0.1375$ &$0.0007$ &$0.0015$ &$0.29$ &$0.171$ &$0.75$ &$0.014$ &$-0.18$ &$0.274$ &$0.32$ &$0.169$ \\
HD\,61986 &$0.1442$ &$0.1393$ &$0.0021$ &$0.0011$ &$0.38$ &$0.004$ &$0.18$ &$0.298$ &$0.27$ &$0.033$ &$-0.30$ &$0.189$ \\
HD\,63765 &$0.2063$ &$0.1975$ &$0.0048$ &$0.0125$ &$0.96$ &$0.001$ &$0.86$ &$0.000$ &$0.85$ &$0.003$ &$0.62$ &$0.001$ \\
HD\,65277 &$0.1962$ &$0.1796$ &$0.0062$ &$0.0047$ &$0.55$ &$0.034$ &$0.61$ &$0.005$ &$-0.50$ &$0.048$ &$0.25$ &$0.153$ \\
HD\,65907 &$0.1352$ &$0.1322$ &$0.0005$ &$0.0006$ &$0.47$ &$0.013$ &$0.23$ &$0.027$ &$0.14$ &$0.248$ &$0.05$ &$0.336$ \\
HD\,68978 &$0.1526$ &$0.1434$ &$0.0042$ &$0.0013$ &$0.89$ &$0.000$ &$0.11$ &$0.364$ &$0.70$ &$0.000$ &$-0.12$ &$0.354$ \\
HD\,69611 &$0.1285$ &$0.1264$ &$0.0007$ &$0.0007$ &$-0.43$ &$0.148$ &$0.14$ &$0.267$ &$0.18$ &$0.332$ &$-0.12$ &$0.304$ \\
HD\,69830 &$0.1499$ &$0.1356$ &$0.0026$ &$0.0010$ &$0.81$ &$0.016$ &$0.26$ &$0.088$ &$-0.07$ &$0.425$ &$-0.36$ &$0.028$ \\
HD\,71334 &$0.1307$ &$0.1306$ &$0.0009$ &$0.0014$ &$0.32$ &$0.215$ &$0.28$ &$0.144$ &$-0.43$ &$0.145$ &$0.06$ &$0.410$ \\
HD\,71835 &$0.1640$ &$0.1346$ &$0.0018$ &$0.0021$ &$-0.30$ &$0.199$ &$0.41$ &$0.042$ &$0.67$ &$0.031$ &$0.18$ &$0.218$ \\
HD\,7199 &$0.1796$ &$0.1283$ &$0.0076$ &$0.0033$ &$0.86$ &$0.018$ &$0.04$ &$0.398$ &$-0.64$ &$0.057$ &$-0.11$ &$0.263$ \\
HD\,72673 &$0.1619$ &$0.1459$ &$0.0025$ &$0.0014$ &$0.71$ &$0.000$ &$0.25$ &$0.239$ &$0.35$ &$0.040$ &$-0.20$ &$0.285$ \\
HD\,73524 &$0.1270$ &$0.1219$ &$0.0037$ &$0.0007$ &$-0.42$ &$0.074$ &$-0.22$ &$0.102$ &$0.10$ &$0.358$ &$0.23$ &$0.093$ \\
HD\,7449 &$0.1478$ &$0.1438$ &$0.0020$ &$0.0017$ &$0.73$ &$0.000$ &$0.68$ &$0.015$ &$0.75$ &$0.000$ &$0.19$ &$0.273$ \\
HD\,77110 &$0.1399$ &$0.1337$ &$0.0011$ &$0.0020$ &$0.14$ &$0.337$ &$-0.09$ &$0.263$ &$0.08$ &$0.408$ &$-0.20$ &$0.080$ \\
HD\,78429 &$0.1464$ &$0.1327$ &$0.0020$ &$0.0047$ &$-0.31$ &$0.206$ &$-0.18$ &$0.332$ &$-0.48$ &$0.105$ &$-0.64$ &$0.060$ \\
HD\,79601 &$0.1318$ &$0.1290$ &$0.0003$ &$0.0015$ &$-0.86$ &$0.018$ &$0.16$ &$0.229$ &$-0.39$ &$0.166$ &$0.19$ &$0.196$ \\
HD\,82342 &$0.1824$ &$0.1670$ &$0.0031$ &$0.0042$ &$-0.41$ &$0.112$ &$0.17$ &$0.233$ &$0.32$ &$0.166$ &$-0.45$ &$0.024$ \\
HD\,82516 &$0.1840$ &$0.1461$ &$0.0067$ &$0.0026$ &$0.67$ &$0.030$ &$-0.34$ &$0.143$ &$0.52$ &$0.071$ &$-0.01$ &$0.489$ \\
HD\,82943 &$0.1370$ &$0.1252$ &$0.0013$ &$0.0009$ &$0.64$ &$0.013$ &$0.38$ &$0.124$ &$-0.02$ &$0.470$ &$0.50$ &$0.066$ \\
HD\,85390 &$0.1583$ &$0.1413$ &$0.0040$ &$0.0024$ &$0.86$ &$0.017$ &$-0.04$ &$0.465$ &$0.57$ &$0.082$ &$-0.54$ &$0.094$ \\
HD\,87838 &$0.1318$ &$0.1279$ &$0.0022$ &$0.0016$ &$0.00$ &$0.500$ &$0.00$ &$0.499$ &$0.50$ &$0.050$ &$0.08$ &$0.322$ \\
HD\,8828 &$0.1355$ &$0.1319$ &$0.0010$ &$0.0012$ &$0.34$ &$0.143$ &$-0.02$ &$0.473$ &$-0.09$ &$0.387$ &$-0.05$ &$0.419$ \\
HD\,88725 &$0.1436$ &$0.1394$ &$0.0007$ &$0.0026$ &$-0.57$ &$0.066$ &$0.12$ &$0.271$ &$-0.43$ &$0.132$ &$-0.03$ &$0.436$ \\
HD\,89839 &$0.1251$ &$0.1192$ &$0.0086$ &$0.0022$ &$0.01$ &$0.488$ &$0.10$ &$0.377$ &$0.06$ &$0.406$ &$-0.10$ &$0.376$ \\
HD\,90156 &$0.1385$ &$0.1375$ &$0.0005$ &$0.0011$ &$-0.43$ &$0.030$ &$-0.15$ &$0.312$ &$0.28$ &$0.116$ &$0.30$ &$0.173$ \\
HD\,92719 &$0.1595$ &$0.1542$ &$0.0023$ &$0.0026$ &$0.85$ &$0.002$ &$0.95$ &$0.006$ &$0.68$ &$0.012$ &$0.74$ &$0.026$ \\
HD\,93083 &$0.1982$ &$0.1405$ &$0.0066$ &$0.0025$ &$0.29$ &$0.131$ &$-0.69$ &$0.035$ &$-0.38$ &$0.071$ &$-0.43$ &$0.129$ \\
HD\,93385 &$0.1273$ &$0.1255$ &$0.0004$ &$0.0014$ &$-0.51$ &$0.040$ &$-0.62$ &$0.031$ &$-0.14$ &$0.310$ &$-0.59$ &$0.039$ \\
HD\,95456 &$0.1257$ &$0.1183$ &$0.0011$ &$0.0008$ &$0.69$ &$0.004$ &$0.06$ &$0.355$ &$-0.51$ &$0.025$ &$0.14$ &$0.207$ \\
HD\,9578 &$0.2157$ &$0.1997$ &$0.0073$ &$0.0071$ &$0.46$ &$0.128$ &$-0.06$ &$0.402$ &$0.75$ &$0.034$ &$-0.13$ &$0.280$ \\
HD\,96423 &$0.1263$ &$0.1247$ &$0.0007$ &$0.0013$ &$0.13$ &$0.328$ &$-0.22$ &$0.245$ &$-0.24$ &$0.209$ &$0.15$ &$0.326$ \\
HD\,967 &$0.1441$ &$0.1418$ &$0.0005$ &$0.0023$ &$0.75$ &$0.013$ &$0.78$ &$0.001$ &$0.62$ &$0.031$ &$0.16$ &$0.255$ \\
HD\,96700 &$0.1346$ &$0.1324$ &$0.0012$ &$0.0011$ &$0.37$ &$0.015$ &$-0.20$ &$0.131$ &$-0.10$ &$0.277$ &$-0.13$ &$0.230$ \\
HD\,97037 &$0.1279$ &$0.1263$ &$0.0008$ &$0.0011$ &$-0.05$ &$0.435$ &$-0.35$ &$0.067$ &$-0.71$ &$0.017$ &$-0.23$ &$0.166$ \\
HD\,97343 &$0.1324$ &$0.1300$ &$0.0010$ &$0.0012$ &$0.52$ &$0.061$ &$0.10$ &$0.349$ &$-0.05$ &$0.435$ &$0.54$ &$0.019$ \\
\end{longtable}
\tablefoot{
$\langle S_\text{CaII} \rangle$ is the mean \sca{}, $\rho^{06}$, $\rho^{16}$ are the correlation coefficients between \sca{} and \nsha{}, and \wsha{}, respectively. $p$-value is the probability of having an equal or higher correlation coefficient (see text). $p$-values of $0.000$ means that the value is lower than $10^{-3}$. The subscripts 'max' and 'min', refer to the short-term correlations at epochs when the star is at maximum and minimum of activity, respectively.
}
}

\end{appendix}

\end{document}